\documentclass{aa}
\usepackage{graphics}

\begin{document}

 
\title{The new sample of Giant radio sources:\\
I. Radio imaging, optical identification and spectroscopy of selected
candidates}
 
\author{J. Machalski, M. Jamrozy, S. Zola}
\offprints{J. Machalski,
\email: machalsk@oa.uj.edu.pl}
\institute{Astronomical Observatory, Jagellonian University,
ul. Orla 171, PL-30244 Cracow, Poland}
\date{Received  / Accepted }
 
\abstract{
A new sample of very large angular size radio sources  has been
selected from the 1.4 GHz VLA surveys: FIRST and NVSS. This sample will be very
useful for an observational constraint on the time evolution of double radio
sources, especially their size, predicted by numerous analytical models of such
evolution (cf. Introduction). In this paper we present radio and optical data
for a large fraction of the sample sources. They are: high-frequency VLA maps
with very weak radio cores detected, deep optical images showing the identified
faint host galaxies not visible on the DSS images, and optical spectra of the
identified galaxies brighter than about $R\approx 18.5$ mag taken with the
McDonald Observatory 2.1m telescope.
For 15 galaxies (of which 4 do not belong to the complete sample) the redshift
has been determined. In the result, 44 per cent of galaxies in the complete
sample have redshift data (with one exception all redshifts are less than
0.33), of which 70 per cent have a linear size exceeding 1 Mpc. The photometric
redshift estimates for other 11 galaxies with 19 mag$<R<$21.7 mag ($0.3<z<0.5$)
predict their sizes much over 1 Mpc.
}
 
\authorrunning{J. Machalski et al.}
\titlerunning{Low-luminosity Giant radio sources}
\maketitle
 
\section{Introduction}
 
Giant double radio sources with a projected linear extent $D\geq 1$ Mpc
(if $\Omega_{o}=1$ and $H_{o}=50$ km~s$^{-1}$Mpc$^{-1}$ which we use
hereafter) are the largest objects in the Universe. Since the existing models
of the radio source evolution (e.g. Kaiser et al. 1997;
Kaiser \& Alexander 1999; Blundell et al. 1999) predict that
classical double sources grow in size with increasing  age, these giants must
represent a very late phase of the evolution of radio sources. In the analytical
model of Kaiser et al. (1997), the range of luminosities and the largest
linear sizes are limited mostly by the jet powers and the different processes
of energy loss. The importance of adiabatic, synchrotron, and inverse Compton
losses in reducing the source luminosity at large life time is so significant
that the largest sources should have very low luminosity, especially at high
observing frequencies. Blundell et al. argue that the samples of FRII sources
are strongly affected by the effect of "youth--redshift degeneracy" causing
the old sources are barely observed at high redshifts.
 
A natural consequence of these predictions is that the majority of giants have
to be very old. On the other hand, it is not
clear under what circumstances radio sources can live long enough to reach 
Mpc-sizes, and what fraction of their population can do that. One 
possibility is that the extreme sizes of some sources are related to their
surrounding intergalactic medium. Mack et al. (1998), after the spectral ageing
analysis of 5 giant radio galaxies, found evidence that those giants are so
large because of their low-density environments and not because of higher ages.
Kaiser \& Alexander suggested that a few giant sources in the LRL sample
(Laing et al. 1983) may "constitute a class of objects intrinsically
different from the rest of the sample".
 
It has to be emphasized that all these studies were based on an inhomogenous
collection of currently known giant sources. An ambitious project to select a
uniform and large sample of giant radio sources appropriate for an extensive
study of their population, has been undertaken at  Leiden Observatory  (e.g.
Schoenmakers et al. 1998). The sample of 33 candidate sources, selected from
the WENSS 325-MHz survey in the sky area of about 2.5 sr (Rengelink et al. 1997),
has been studied by Schoenmakers (1999). These sources supplement a sample of 19
Giants already known in the same sky area.
However, in our contribution during IAU Symposium No. 199 on the population
modelling of giant FRII-type sources, we have presented evidence that only a
very small fraction of expected faint giants (with 1.4 GHz flux density less
than 0.5 Jy) has already been detected. In order to find the missing giant sources
we have selected another sample of faint candidates for which  further radio and
optical observations have been done.
 
In this paper, the first results of (i) the radio imaging with the main effort
to detect a compact radio core in those giant candidates whose cores were not
detected yet, (ii) the follow-up optical identifications, and (iii) the optical
spectroscopy of identified optical counterparts (with the one exception - only
galaxies) are presented. In Sect.~2, the selection criteria and the list of
candidates are given. The new radio observations are described in Sect.~3. The
optical identification, and the deep optical imaging for those candidates whose
optical counterpart is not visible on the POSS plates, are gived in Sect.~4.
Finally, the optical spectroscopy of the identified galaxies bright enough that
their spectrum could be taken with a medium-size telescope are presented in
Sect.~5.
 
Basic physical parameters have been derived for the sample sources
with spectroscopic redshift including their radio and optical luminosity,
projected linear size, equipartition magnetic field and energy density. A
photometric redshift and size estimates are provided for a few host galaxies
identified in the deep optical images and for which a crude apparent $R$-band
magnitude has been determined. These results are summarized in Sect.~6.
 
\section{Selection criteria}
 
We decided to select our candidates from the 1.4-GHz sky surveys: NVSS (Condon
et al. 1998) and the first part of FIRST (Becker et al. 1995). These
surveys, which provide radio maps made with two different angular resolutions
($45\arcsec$ and $5\arcsec$, respectively) at the same observing frequency,
allowed (i) an effective removal of confusing sources, (ii) a reliable
determination of its morphological type, and (iii) a determination of the compact
core component necessary for the proper identification of the source with its
host optical object. From the population modelling (cf. Introduction) we
expected that most  missing giants should be below a redshift of about 0.4.
At $z\approx 0.4$, a source with a linear size of 1 Mpc will have an angular
size of 2.6 arc min. Therefore, we have chosen the following selection criteria:

The radio source
 
\noindent
(1) lies within the sky area of 0.47 sr limited by
$07^{\rm h}20^{\rm m}<{\rm R.A.(J2000)}<17^{\rm h}30^{\rm m}$ and
$+28\fdg 5<{\rm Decl.(J2000)}<+41\fdg 0$,
 
\noindent
(2) has a morphological type (suggested from the NVSS and FIRST radio maps, Fanaroff-Riley (1974)) type II (FRII) or FRI/II,
 
\noindent
(3) has an angular separation between the brightest regions on the maps
$\Theta\stackrel{>}{\sim}3$ arc min, and
 
\noindent
(4) has a 1.4 GHz flux density on the NVSS map of $S_{1.4}<500$ mJy.
 
The above criteria were fulfilled by 36 sources listed in Table~1. The entries of
Table~1 are:
 
{\sl Column 1:} IAU name at epoch J2000
 
{\sl Column 2:} Fanaroff-Riley morphological type
 
{\sl Column 3:} 1.4 GHz flux density from the NVSS survey
 
{\sl Column 4:} radio spectral index around the frequency of 1.4 GHz specified 
as the mean of spectral indices $\alpha(325,1400)$ and $\alpha(1400,4850)$. If 
a flux density at about 5 GHz is not known, the respective spectral index is given 
as $>\alpha(325,1400)$
 
{\sl Column 5:} angular size in arc sec
 
{\sl Column 6:} information about radio core; `c' indicates that the core has already been
detected in the FIRST survey, \underline{\bf `c'} -- that it has been detected in this work.
`c?' indicates the lack of high-resolution map, but a compact core is very
probable
 
{\sl Column 7:} optical identification; G--galaxy, Q--quasar
 
{\sl Column 8:} R--band apparent magnitude of the identified host object taken 
from the reference given in column 10. The magnitudes of faint galaxies detected 
in this paper (cf. Sect.~4.2 and Table~4) are marked by bold-face
 
{\sl Column 9:} redshift of the host object (galaxy). The redshifts of 5 objects
were available prior to our spectroscopical observations; the relevant references
are given in column 10. Redshifts determined in this paper are marked by
bold-face.
 
{\sl Column 10:} references to optical type of the identified object and its
magnitude, and to the redshift.
 
At the bottom of Table 1 we list 4 additional giant candidates which have been
observed spectroscopically though they do not belong to the complete sample.
J1218+5026, taken from the GB/GB2 sample (Machalski 1998), was supposed to be a
very large source; the remaining three (J1018--1240, J1328--0307, and J1457--0613)
are the largest sources among the Las Campanas radio galaxies (cf. Machalski \&
Condon 1999).

{\renewcommand{\baselinestretch}{1.3} 
\begin{table*}
\caption{The sample of Giant candidates}
\begin{tabular*}{140mm}{llrlcclllll}
\hline
IAU name & FR & $S_{1.4}$ & $\alpha$ &$\Theta$& radio & opt. & R & z & Ref & Notes\\
J(2000)&type & [mJy]  &    &[\arcsec] & core  & id.  & [mag] \\
\hline
J0720+2837 & II   &   46 &$>$0.7 & 370 & c  & G & 17.75 &{\bf 0.2705} & 1,6\\
J0725+3025 & II   &   35 & 1.43  & 175 & c  &...& ..... & ..... \\
J0816+3347 & II   &   40 & 1.15  & 210 & c  & Q?& 18.8-{\bf 20.7}&.....& 1,6 & +\\
J0912+3510 & II   &  143 & 1.24  & 375 &... & G & 19.35 &{\bf 0.2489} & 1,6 & +\\
           &      &      &       &     &    & G & 19.5  &           & 6\\
J0927+3510 & II   &   96 & 1.20  & 345 &... & 2G&{\bf 20.6}or{\bf 21.7}&.....& 6 & +\\
J1011+3111 & II   &   70 &$>$0.6 & 286 & c  & G &{\bf 21.2} & .....  & 6\\
J1113+4017 & I/II &  246 & 0.74  & 720 & c  & G & 14.56 & 0.0745 & 1,4\\
J1155+4029 & II   &  323 & 1.04  & 229 & c  & G &{\bf 21.5}  & .....  & 6\\
J1200+3449 & II   &  223 & 1.05  & 147 &\underline{{\bf c}}& G &{\bf 21.2}& ..... & 6\\
J1253+4041 & I/II &   52 &$>$0.8 & 275 &... & G & 17.27 &{\bf 0.2302} & 1,6\\
J1254+2933 & II   &   65 & (0.6) & 300 &\underline{{\bf c}}& G &{\bf 20.3}& ..... & 6\\
J1330+3850 & II?  &   31 & 0.77  & 280 &... &2G &{\bf 19.3}or{\bf 19.6}&.....& 6 & +\\
J1343+3758 & II   &  136 & 0.87  & 678 &\underline{{\bf c}}& G & 17.94 &{\bf 0.2267} & 1,6&+\\
J1344+4028 & I/II &  222 & 0.78  & 450 & c  & G & 14.92 &{\bf 0.0748} & 1,6 &+\\
J1345+3952 & II   &  170 & 1.14  & 167 &... & G & 15.80 &{\bf 0.1611} & 1,6\\
J1355+2923 & II   &  140 & 1.27  & 263 & c  & G &{\bf 20.4} & .....  & 6\\
J1428+2918 & II   &  431 & 1.10  & 905 & c  & G & 13.0  & 0.0870 & 4\\
J1428+3938 & II   &   89 & 1.09  & 230 & c  &...& ..... & .....  \\
J1445+3051 & II   &   95 & 0.91  & 290 & c  & G & 19.01 & 0.42   & 3\\
J1451+3357 & II   &  138 & 0.91  & 245 & c  & G & 18.7  &{\bf 0.3251} & 6\\
J1453+3309 & II   &  460 & 0.96  & 320 & c  & G & 18.3  & 0.249  & 4\\
J1512+3050 & II   &   98 & 0.67  & 240 & c  & G & 15.98 &{\bf 0.0895} & 1,6\\
J1513+3841 & II?  &   25 &$>$0.6 & 210 &... &...& ..... & .....  \\
J1525+3345 & II   &   51 &$>$0.8 & 214 & c  & G &{\bf 20.9} & .....  & 6\\
J1526+3956 & II   &   62 &$>$1.0 & 536 & c? &...& ..... & .....  \\
J1554+3945 & II   &   74 &$>$0.8 & 219 &\underline{{\bf c}}& G & 19.5  & ..... & 6\\
J1555+3653 & II   &  106 & 1.00  & 335 & c  & G & 18.56 &{\bf 0.2472} & 1,6 &+\\
J1604+3438 & II   &  146 & 0.95  & 200 & c? &...& ..... & .....  \\
J1604+3731 & II   &  122 & 1.05  & 182 & c  & G?& ..... & 0.814  & 2\\
J1615+3826 & II?  &   37 & (0.5) & 245 &... & G & 17.59 &{\bf 0.1853} & 1,6\\
J1632+3433 & II   &   29 &$>$0.9 & 180 & c  &...& ..... & .....  \\
J1635+3608 & I/II &  100 & 0.63  & 320 & c  & G & 17.30 &{\bf 0.1655} & 1,6\\
J1649+3114 & II   &  153 & 1.10  & 209 & c  & G & 19.62 & .....  & 1\\
J1651+3209 & I/II &   68 & (0.5) & 430 & c  &...& ..... & .....  \\
J1712+3558 & II   &   86 & 0.93  & 210 &\underline{\bf c}& G & 19.1  & ..... & 6\\
J1725+3923 & II   &   88 & 0.95  & 270 & c  & G & 18.76 & .....  & 1 &+\\
\hline
& &\\
J1018--1240 & I/II & 276 & 1.27  & 555 & c  & G & 16.36 &{\bf 0.0777} & 5,6\\
J1218+5026  & II   & 596 & 0.78  & 196 & c? & G & 17.8  &{\bf 0.1995} & 6\\
J1328--0307 & II   & 201 &$>$1.2 & 801 & c  & G & 16.82 &{\bf 0.0860} & 5,6\\
J1457--0613 & II   & 507 &$>$1.0 & 235 & c? & G & 17.29 &{\bf 0.1671} & 5,6\\
\hline
\end{tabular*}
 
References: (1) the Digitized Sky Survey (DSS) data base, (2) Cotter et al.\,(1996),\\
(3) Hook et al.\,(1998), (4) Schoenmakers (1999), (5) Machalski \& Condon (1999),\\
(6) this paper
\end{table*}
 }
{\bf Notes on individual sources in Table 1:}
 
{\sl J0816+3347:} Very blue object on the DSS (E=18.76 mag; O--E$=-0.04$ mag).
R=($20.7\pm 0.2$) mag was found in March 4, 2000  (cf. Fig.~3a). A very 
certain variability suggests that this object may be a quasar.
 
{\sl J0912+3510:} The source was noticed as a giant by Jamrozy \& Machalski
(1999) on the basis of a photometric redshift estimate derived for the brightest
galaxy near the midpoint between its radio lobes (marked as galaxy A in Table 4 
and in Fig.~2a).
Fig.~4b shows the spectrum of galaxy A. On the other hand, the statistical
effect that the radio core is predominantly closer to the brighter radio lobe
(cf. Sect.~6.4) suggests another possible identification with galaxy B.
Unfortunately, no radio core brighter than about 0.1 mJy\,beam$^{-1}$ was
detected in spite of two observational attempts (cf. Table~2).
 
{\sl J0927+3510:} The second case where we did not detect the radio core 
 (cf. Fig.~1a). Therefore,
a probable identification is a galaxy marked as galaxy A in Table~4 and Fig.~3b.
This 20.6 mag galaxy is closer to the brighter eastern lobe of the source, thus
is favoured by the correlation mentioned above. The second possible
identification is the 21.7 mag galaxy B lying close to the  midpoint between the radio
lobes (cf. Fig.~3b).
 
{\sl J1330+3850:} Because a radio core is unknown yet, the optical identification
remains uncertain. Two possible identifications of a host galaxy are given in
Table~4.
 
{\sl {\bf J}1343+3758:} The radio map with the core indicated is shown in 
Fig.~1d. A detailed analysis of this the third largest giant source
known up to now is given by Machalski \& Jamrozy (2000). The final corrected
redshift of the identified galaxy is given in this paper.
 
{\sl J1344+4028:} A 20 mJy background source at $13^{\rm h}44^{\rm m}46\fs 51$,
$+40\degr 26\arcmin 50\farcs 6$ (J2000), contaminating the sample source, is
subtracted.
 
{\sl J1555+3653:} The sample source is slightly confused on the NVSS map 
by the strong steep-spectrum background source 6C155304.9+370235 (B1950).
 
{\sl J1725+3923:} Strongly confused by two compact sources at the (J2000)
positions: $17^{\rm h}25^{\rm m}18\fs 15$; $+39\degr 21\arcmin 19\farcs 0$ and
$17^{\rm h}25^{\rm m}24\fs 44$; $+39\degr 24\arcmin 04\farcs 7$, hence its
radio spectrum is uncertain.

\begin{table*}
\caption{The VLA observing log}
\begin{tabular*}{109mm}{lccccl}
\hline
Source    &$\nu$ & beam size  & int. time  & rms noise  & observing\\
          & [GHz]&[$\arcsec\times\arcsec$]& [min] & [$\mu$Jy/beam] & date\\
\hline
J0912+3510 & 4.86 & $1.3\times 1.0$&$3\times 11$ & 50 & May\,24,1997\\
          & 8.46 & $3.5\times 0.9$ &$2\times 25$ & 21 & Feb.20,2000\\
J0927+3510 & 8.46 & $2.8\times 0.9$ &$2\times 26$ & 23 & Feb.20,2000\\
J1200+3449 & 8.46 & $2.4\times 0.8$ &$2\times 23$ & 44 & Feb.20,2000\\
J1254+2933 & 4.86 & $1.3\times 1.2$ &$3\times 23$ & 33 & Dec.13,1999\\
J1343+3758 & 4.86 & $1.2\times 1.1$ &$3\times 24$ & 30 & Dec.13,1999\\
J1513+3841 & 4.86 & $3.0\times 1.1$ &$3\times 22$ & 52 & Feb.12,2000\\
J1554+3945 & 4.86 & $3.0\times 1.2$ &$3\times 22$ & 43 & Feb.12,2000\\
J1712+3558 & 4.86 & $3.2\times 1.1$ &$3\times 22$ & 55 & Feb.12,2000\\
\hline
\end{tabular*}
\end{table*}

\begin{table*}
\caption{Data of the detected radio cores}
\begin{tabular*}{109mm}{lcllcc}
\hline
Source     &$\nu$ & R.A.(J2000) & Dec.(J2000) & $S_{\rm core}$ & $f_{\rm c}$\\
           &[GHz] &             &             & [mJy]\\
\hline
J1200+3449  & 8.46 &$12^{\rm h}00^{\rm m}50\fs 25$ &
$+34\degr 49\arcmin 20\farcs 6$ & $0.36\pm 0.08$  & 0.002\\
J1254+2933  & 4.86 & 12 54 34.06 & +29 33 40.2 & $0.74\pm 0.09$ & 0.011\\
J1343+3758  & 4.86 & 13 42 54.53 & +37 58 18.8 & $1.09\pm 0.07$ & 0.008\\
J1554+3945  & 4.86 & 15 54 26.92 & +39 45 08.7 & $0.84\pm 0.09$ & 0.011\\
J1712+3558  & 4.86 & 17 12 24.89 & +35 58 26.2 & $0.55\pm 0.12$ & 0.006\\
\hline
\end{tabular*}
\end{table*}

\section{Radio observations}

Eight out of twelve sources with no radio core detected in the FIRST survey have
been observed with the VLA in the B--array at 4.86 GHz and/or BnC--array at
8.46 GHz. The observing log is given in Table~2. In order to reach a $rms$
noise of about 30\,$\mu$Jy\,beam$^{-1}$ the field of view centred at a midpoint
between the lobes  was observed for an integration time of at least 40 min. 
The interferometric phases were calibrated approximately every 20 min with the 
phase calibrator nearest to the observed source. The source 3C286 served as the
primary flux density calibrator.

The sky coordinates and the integrated flux density  of the detected cores are
given in Table~3. The last column of Table 3 gives the ratio of 5-GHz flux
density of the core to 1.4-GHz flux density of the entire source. Unfortunately,
no core brighter than about 0.1 mJy\,beam$^{-1}$ was detected in the sources
J0912+3510 and J0927+3510. Dividing the above limit of core flux by the total
flux
of these sources, one can expect their fraction $f_{\rm c}$ to be below 0.001.
This is about 2 times less than that of the faintest core detected in already
known giants. Also no core was detected in the source J1513+3841. In this 
case, the upper limit of 5 GHz flux density of the core is about 0.25 mJy beam$^{-1}$.

The high-frequency VLA contour maps for those sources whose extended lobes were
not attenuated too much by the VLA primary beam and/or where some compact structures
reside in the lobes, are presented in Figs.~1a--f. In order to show how these
structures are located inside the extended emission, each contour map is
overlaid onto the relevant NVSS map shown in the gray scale. The compact cores, for
the sake of proper contrast, are shown by the black contours. The crosses
indicate the position of the host galaxy (cf. Sect.~4).
 
\section{Optical imaging and identification}
 
\subsection{DSS identification}
 
To identify our source candidates whose radio core is known with a host optical
object, first of all we used the `Digitized Sky Survey' (DSS), i.e. digitized
POSS survey. For 29 sample sources with cores, we have identified their cores
with 13 galaxies and one very blue object, possibly a quasar. This means that
the  optical counterpart of one half of our giant candidates is fainter than
$R\approx 20$ mag. Taking into account the $R$-band Hubble diagram for 3C radio
galaxies (Eales 1985) and allowing the giant radio galaxies to be one magnitude
less luminous than the 3C ones, $R\stackrel{>}{\sim}20$ mag implied a redshift
$z>0.4$. For four giant candidates without  known radio core (J0912+3510, 
J1253+4041,
J1345+3952, and J1615+3826) a very probable host galaxy has been determined.
The optical fields, reproduced from the DSS and showing the presumed
identification with the radio contour NVSS map overlayed, are shown in
Fig.~2a--d. The $R$ magnitudes of the identified galaxies from the DSS are listed
in Table~1 with 0.01 precision.

\subsection{Deep imaging}
 
For 9 sky fields containing the sample sources, 6 of which having no radio core
identification in the DSS,  deep optical imaging was done  with the
2.1m telescope of the McDonald Observatory (Texas). The observations were made
using the `Imaging Grism Instrument' (IGI) equipped with aTK4
$1024\times 1024$ CCD detector cooled with liquid nitrogen. IGI allows direct
imaging and spectroscopy with a spatial scale of $0\farcs 48$ per pixel within
the field of view of about $8\times 8$ arc min. The sky fields centred {\bf on} the
radio source were observed through the Cousins $R$-band filter at zenith angles
providing air masses less than 1.2, most of them being below 1.05. The exposure
times applied ranged from 3 min to 10 min. 
Although seeing conditions during the
observing night of March 4/5,2000  were not photometric ones, also the standard
fields NGC\,2419, NGC\,4147, and M92 (Christian et al., 1985) were observed for
a crude photometric calibration.
 
The astrometric calibration has been done by transforming the instrumental
pixel coordinates of stars in the investigated frame into their sky coordinates
in the DSS data base. As a result, all six sources with a core have been
identified with faint galaxies. In the two remaining fields, possible galaxy
identifications are found. Instrumental magnitudes of the identified host
galaxies (and several other objects) on the CCD frames, reduced for bias, dark
current and flat field with the ESO MIDAS package, were determined using both
the aperture (MIDAS magnitude/circle procedure) and PSF (DAOPHOT II procedure;
Stetson 1987) methods. These instrumental magnitudes were then transformed into
$R$ magnitudes in the twofold manner: a transformation formula was determined
for (i) those objects in the particular frame whose magnitudes are calibrated in
the DSS ($R_{\rm DSS}$), and (ii) standards in the calibration fields
($R_{\rm cal}$). Dispersions of the differences $\Delta R_{\rm 1}=R-R_{\rm DSS}$
and $\Delta R_{\rm 2}=R-R_{\rm cal}$ (the latter calculated from the calibration
fields taken just before and after the target frame) have been used to determine
the combined $rms$ error
$\Delta R=(\Delta R_{\rm 1}^{2}+\Delta R_{\rm 2}^{2})^{1/2}$ of particular $R$
magnitude.

The sky coordinates, radio-minus-optical offsets with respect to the radio core,
and $R$-magnitude estimate with its $rms$ error of the identified faint galaxies
are listed in Table~4. The CCD R-band frames with the deep identification marked
are shown in Figs.~3a--h.
 
\begin{table*}
\caption{Data of the identified faint galaxies}
\begin{tabular*}{120mm}{lcllccc}
\hline
Source & gal. & R.A.(J2000) & Dec.(J2000) & $\Delta$R.A. & $\Delta$Dec. & R$\pm\Delta$R\\
       &  &             &             & [s]          & [\arcsec]    & [mag]\\
\hline
J0912+3510&A&$09^{\rm h}12^{\rm m}51\fs 81$&$+35\degr 10\arcmin 16\farcs 1$&
..... & ..... & 19.35\\
      & B & 09 12 51.20 & +35 10 07.6 & .....   & .....  & $19.5\pm 0.2$\\
J0927+3510&A&09 27 54.93 & +35 10 39.8 & .....  & .....   & $20.6\pm 0.3$\\
      & B & 09 27 50.58 & +35 10 50.5 & .....   & .....  & $21.7\pm 0.4$\\
J1011+3111& & 10 11 12.21 & +31 11 05.1 & $-0.07$ & $+0.1$ & $21.2\pm 0.3$\\
J1155+4029& & 11 55 49.61 & +40 29 40.8 & $-0.06$ & $-0.2$ & $21.5\pm 0.4$\\
J1200+3449& & 12 00 50.56 & +34 49 21.0 & $-0.28$ & $-0.4$ & $21.2\pm 0.4$\\
J1254+2933& & 12 54 34.14 & +29 33 41.8 & $-0.08$ & $-1.6$ & $20.3\pm 0.2$\\
J1330+3850&A& 13 30 36.33 & +38 50 20.0 & .....   & .....  & $19.3\pm 0.2$\\
          &B& 13 30 34.84 & +38 50 19.9 & .....   & .....  & $19.6\pm 0.3$\\
J1355+2923& & 13 55 17.83 & +29 23 34.5 & $-0.18$ & $-0.6$ & $20.4\pm 0.2$\\
J1525+3345& & 15 25 00.84 & +33 45 42.9 & $-0.06$ & $-1.2$ & $20.9\pm 0.3$\\
\hline
\end{tabular*}
\end{table*}

\section{Optical spectroscopy}
 
\subsection{Observations}
 
In order to determine redshifts of the identified galaxies brighter than about
18.5 mag in the $R$ band, optical spectroscopic observations were conducted
between Feb 27 and March 4, 2000. Also these observations were made with the
McDonald Observatory 2.1m telescope equipped with the IGI instrument. A grism
sensitive to the wavelength range of 3750 {\AA} to 7600 \AA, and a $2\arcsec$ wide
slit providing a dispersion of 3.7 {\AA} per pixel and spectral resolution of
about 12 \AA, were used. A number of exposures ranging from 15 min to 30 min was
taken for 15 galaxies, of which only 11 belonging to the
complete sample were bright enough to be observed with the 2.1m telescope.
The resulting redshifts have been put in bold face in Table~1. The epoch of observations did not allow us to try to observe 
the identified galaxies at R.A.$>16^{\rm h}30^{\rm m}$ (i.e. J1712+3558 and J1725+3923).
 
Usually two or three useable exposures were obtained for each galaxy, which
made it possible to improve S/N ratio without an increase of smearing of the spectrum
during a too long single exposure due to imperfect tracking of the telescope.
The total integration time for each spectrum ranged from 30 min for the 
brightest galaxy (J1344+4028) to 90 min for the faintest ones.
The wavelength calibration was carried out using exposures to helium and mercury
lamps. The flux calibration was provided by exposures of a spectrophotometric
standard star close to the observed galaxy, chosen from among HZ43, HZ44, EG50,
EG79, GZ140, and Feige98 (Stone 1977).
 
\subsection{Data reduction and the results}
 
All spectra obtained were reduced in the standard way, i.e. corrected for bias,
dark current and flat field, rejected cosmic rays , combined and calibrated using the
{\sc longslit} package of the IRAF software. One-dimensional spectra for the
galaxies were extracted interactively, using the observed intensity profile
along the slit to define the apertures and sky-background regions. The final
1--D spectra
of the observed galaxies are shown in Figs.~4a--o. The emission lines and/or
absorption bands detected, as well as the resultant redshift of these features
with its $rms$ error are listed in Table~5.

{\renewcommand{\baselinestretch}{1} 
\begin{table*}
\caption{Lines detected and redshift of the galaxies}
\begin{tabular*}{180mm}{llllllll}
\hline
Source     & Line/absorp.band & $\lambda_{o}$ & z  & Source  & Line/absorp.band &
$\lambda_{o}$ & z\\
        & detected & [\AA]        &    &        & detected & [\AA]\\
\hline
J0720+2837 & [OII]3727 & 4736.7 & 0.2709 & J1555+3653 & [OII]3727 & 4649.8 & 0.2476\\
           & CaII 3934 & 4998.5 & 0.2706 &            & CaII 3968 & 4947.7 & 0.2469\\
           & CaII 3968 & 5039.1 & 0.2699 &            & G band    & 5365.  & 0.2462\\
           & G band    & 5468.  & 0.2702 &            & [OIII]5007& 6247.0 & 0.2477\\
           & [OIII]4959& 6101.5 & 0.2707 &            & Mg band   & 6456.  & 0.2475\\
           & [OIII]5007& 6361.6 & 0.2705 &  &  &  &$\overline{0.2472}\pm 0.0007$\\
           & Mg band   & 6577.  & 0.2709 \\
      & & & $\overline{0.2705}\pm 0.0006$& J1615+3826 & [OII]3727 & 4416.3 & 0.1849\\
           &           &        &        &            & CaII 3968 & 4703.6 & 0.1854\\
J0912+3510 & [OII]3727 & 4655.2 & 0.2491 &            & [OIII]4959& 5877.3 & 0.1852\\
           & H$\beta$  & 6071.3 & 0.2489 &            & [OIII]5007& 5936.7 & 0.1857\\
           & [OIII]4959& 6192.5 & 0.2487 &  &  &  &$\overline{0.1853}\pm0.0005$\\
           & [OIII]5007& 6252.4 & 0.2487 & \\
      & & & $\overline{0.2489}\pm 0.0004$& J1635+3608 & CaII 3934 & 4590.3 & 0.1668\\
           &           &        &        &            & CaII 3968 & 4627.9 & 0.1663\\
J1253+4041 & [OII]3727 & 4586.2 & 0.2305 &            & G band    & (5014) &(0.165)\\
           & CaII 3934 & 4840.0 & 0.2303 &            & Mg band   & (6020) &(0.163)\\
           & CaII 3968 & 4881.8 & 0.2303 &            & Na\,D     & 6875.  & 0.1666\\
           & G band    & 5295.  & 0.2300 &  &  &  &$\overline{0.1655}\pm 0.0015$\\
           & [OIII]4959& 6098.7 & 0.2298 \\
           & [OIII]5007& 6158.3 & 0.2299 & J1018--1240& CaII 3934 & 4237.8 & 0.0772\\
           & Mg band   & 6367.  & 0.2303 &            & CaII 3968 & 4278.6 & 0.0783\\
           & Na\,D     & 7248.  & 0.2299 &            & G band    & 4641.  & 0.0781\\
      & & & $\overline{0.2302}\pm 0.0005$&            & Mg band   & 5580.  & 0.0783\\
           &           &        &        &            & Na\,D     & 6350.  & 0.0775\\
J1343+3758 & [OIII]4959& 6083.1 & 0.2267 &            & H$\alpha$ & 7065.9 & 0.0768\\
           & [OIII]5007& 6141.7 & 0.2266 &  &  &  &$\overline{0.0777}\pm 0.0007$\\
      & & & $\overline{0.2267}\pm 0.0005$ \\
           &           &        &        & J1218+5026 & [OII]3727 & 4470.4 & 0.1995\\
J1344+4028 & CaII 3934 & 4223.3 & 0.0735 &            & [NeIII]3869 & 4640.9 & 0.1995\\
           & CaII 3968 & 4257.7 & 0.0730 &         &H$\gamma$+[OIII]& 5205.  & 0.199\\
           & G band    & 4620.  & 0.0732 &            & H$\beta$  & 5831.7 & 0.1997\\
           & Mg band   & 5580.  & 0.0783 &            & [OIII]4959& 5948.5 & 0.1995\\
           & Na\,D     & 6340.  & 0.0759 &            & [OIII]5007& 6005.8 & 0.1995\\
      & & & $\overline{0.075}\pm 0.002$   &  &  &  &$\overline{0.1995}\pm 0.0004$\\
      &\\
J1345+3952 & CaII 3934 & 4567.7 & 0.1611 & J1328--0307& [OIII]4959& 5387.1 & 0.0863\\
           & CaII 3968 & 4607.9 & 0.1613 &            & [OIII]5007& 5437.6 & 0.0860\\
           & G band    & 5000.  & 0.1614 &            & H$\alpha$ & 7125.3 & 0.0858\\
           & Mg band   & 6003.  & 0.1601 &  &  &  &$\overline{0.0860}\pm 0.0005$\\
           & Na\,D     & 6846.  & 0.1617 \\
      & & & $\overline{0.1611}\pm 0.0008$& J1457--0613& [OII]3727 & 4350.2 & 0.1672\\
           &           &        &        &            & H$\beta$  & 5674.4 & 0.1673\\
J1451+3357 & [OII]3727 & 4940.7 & 0.3257 &            & [OIII]4959& 5787.2 & 0.1670\\
           & H$\beta$  & 6441.5 & 0.3251 &            & [OIII]5007& 5842.7 & 0.1669\\
           & [OIII]4959& 6599.9 & 0.3248 &  &  &  &$\overline{0.1671}\pm 0.0002$\\
           & [OIII]5007& 6632.9 & 0.3247 \\
      & & & $\overline{0.3251}\pm 0.0005$\\
           & \\
J1512+3050 & CaII 3934 & 4284.8 & 0.0892\\
           & G band    & 4690.  & 0.0894\\
           & Mg band   & 5640.  & 0.0899\\
           & Na\,D     & 6421.  & 0.0896\\
     & & &  $\overline{0.0895}\pm 0.0006$\\
\hline
\end{tabular*}
\end{table*}
 }
Most of the spectra are typical of elliptical and lenticular (E, S0) galaxies
whose continuum emission with the prominent 4000 {\AA} discontinuity is dominated
by evolved giant stars (cf. Kennicutt 1992). The emission lines mostly detected
are [OII]$\lambda 3727$ and [OIII]$\lambda4959$ and $\lambda5007$. Also weak
Balmer lines  are present in some spectra, however usually only H$\beta$ is
detected because the majority of the spectra is redshifted by more than 1.1, so
that in those cases the H$\alpha$ line was beyond the wavelength range observed.
In a few spectra the blue continuum is strong and suggests the presence of
a young population of stars characteristic for early-type spirals.  However 
these blue galaxies are too faint to be morphologically classified on the 
digitized POSS plates. For four galaxies (J1344+4028,
J1345+3952, J1512+3050, and J1635+3608) their redshift can be determined from
the absorption lines CaII$\lambda 3934$ $\lambda 3968$ and bands G, Mg, and
Na\,D only.

\section{Physical parameters}
 
\subsection{Determination of the parameters}
 
The global physical parameters estimated for the sources with spectroscopic
redshift are given in Table 6. The twelve columns of Table 6 give:
 
{\sl Column 1:} IAU name of source
 
{\sl Column 2:} Logarithm of radio luminosity at the emitted frequency of 1.4
GHz ($\log L_{\rm 1.4}$) calculated with the luminosity-distance modulus for
$q_{o}$=0.5, i.e.
 
\[A(z)=2(1+z)-\sqrt{1+z}\]
 
{\sl Column 3:} Absolute $R$-band magnitude of the host galaxy calculated
assuming it is an elliptical with an  optical spectral index
($\alpha_{\rm opt}$)
corresponding to the $B-R$ colour and evolving with redshift accordingly the
`c'-model of Bruzual (1983). Thus
 
\[M_{R}=R-5\log A(z)-2.5\{\alpha_{opt}(z)-1\}\log(1+z)-43.891\]
 
No correction for the foreground extinction has been applied because most of the
galaxies have a galactic latitude $\mid b\mid>50\degr$.
 
{\sl Column 4:} Logarithm of the radio--optical luminosity ratio parameter
defined by
 
\[\log r=\log L(1.4{\rm GHz})-\log L_{\rm opt}({\rm R band}), {\rm where}\]
 
\[\log L_{\rm opt}[{\rm W\,Hz^{-1}}]=-0.4\,M_{R}+13.522\]
 
{\sl Column 5:} Linear size of source ($D$), i.e. projected separation between
the hotspots or the brightest regions in opposite radio lobes
 
{\sl Column 6:} Average magnetic field ($B_{\rm me}$) calculated under
assumption of energy equipartition, a cylindrical geometry of the extended
emission with the base diameter set equal to the average width of the lobes,
usually measured half-way between the core and their brightest regions, using
the prescription of Leahy \& Williams (1984), a filling factor of unity, and
equal distribution of kinetic energy between relativistic electrons and protons.
The total radio luminosity is integrated from 10 MHz to 10 GHz.
 
{\sl Column 7:} Minimum energy density ($u_{\rm me}$) under the same assumptions
 
{\sl Column 8:} Ratio of the equivalent magnetic field of the microwave
background radiation $B_{\rm iC}=0.324(1+z)^{2}$[nT] to the equipartition 
magnetic field in the source $B_{\rm me}$
 
{\sl Column 9:} Ratio of the energy losses by synchrotron radiation to the total
energy losses due to the synchrotron and inverse Compton processes,
$B^{2}_{\rm me}/(B^{2}_{\rm iC}+B^{2}_{\rm me})$
 
{\sl Column 10:} Ratio of the 1.4 GHz flux densities in two opposite radio lobes,
$S_{\rm 1}/S_{\rm 2}$
 
{\sl Column 11:} Ratio of the radio core separation from the hotspot (or the
brightest region) in the brighter lobe to that from the darker one,
$d_{\rm 2}/d_{\rm 1}$
 
{\sl Column 12:} Misalignment angle ($\Delta$) between straight lines connecting
the core with the brightest regions in opposite lobes
 
\begin{table*}
\caption{Physical parameters of the sources with spectroscopic redshift}
\begin{tabular*}{167mm}{llccccccc ccr}
\hline
Source    & log$L_{1.4}$ & $M_{\rm R}$ & log $r$ & $D$ & $B_{\rm me}$ & $u_{\rm me}\,10^{-13}$ &
$\frac{B_{iC}}{B_{me}}$ & $\frac{B^{2}_{me}}{B^{2}_{iC}+B^{2}_{me}}$ &
$S_{1}/S_{2}$ & $d_{2}/d_{1}$ & $\Delta$\\
  & [W/Hz] & [mag] &   & [Mpc] & [nT] & [erg/cm$^{3}$] & & & & & [\degr]\\
\hline
J0720+2837 & 25.19 &--24.4 & 1.9  & 1.91 & & & & & 0.9 & 1.38 & 2\\
J0912+3510 & 25.65 &--22.5 & 3.1  & 1.84 & 0.112 & 1.16 & 4.5 & 0.047 & (0.6)&(1.60)&(2)\\
           &       & ..... & .... & .....& & & & &(1.8)&(1.35)&(1)\\
J1113+4017 & 24.78 &--23.9 & 1.7  & 1.38 & 0.166 & 2.56 & 2.2 & 0.165 & 1.6 & .....&...\\
J1218+5026 & 26.03 &--23.4 & 3.2  & 0.81 & 0.356 &11.80 & 1.3 & 0.370 & 1.3 & 0.71\\
J1253+4041 & 25.11 &--24.3 & 1.9  & 1.28 & 0.086 & 0.69 & 5.7 & 0.030 & 1.7 & 1.10 & 1\\
J1343+3758 & 25.52 &--23.6 & 2.6  & 3.14 & 0.083 & 0.63 & 5.9 & 0.028 & 1.5 & 1.34 & 3\\
J1344+4028 & 24.74 &--23.6 & 1.8  & 0.86 \\
J1345+3952 & 25.32 &--24.7 & 1.9  & 0.60 & 0.190 & 3.33 & 2.3 & 0.159 & 1.2 & 1.24 & 5\\
J1428+2918 & 25.16 &--25.8 & 1.3  & 1.98 &       &      &     &       &     & 0.86 & 5\\
J1445+3051 & 25.90 &--25.0 & 2.4  & 1.91 & 0.148 & 2.05 & 4.4 & 0.049 & 1.6 & 1.35 & 0\\
J1451+3357 & 25.84 &--24.2 & 2.6  & 1.41 & 0.145 & 1.95 & 3.9 & 0.061 & 1.2 & 1.22 & 4\\
J1453+3309 & 26.14 &--23.6 & 3.2  & 1.57 & 0.197 & 3.59 & 2.6 & 0.132 & 1.6 & 1.65 & 12\\
J1512+3050 & 24.60 &--22.9 & 1.9  & 0.54 &       &      &     &       & 1.1 & 1.10 & 5\\
J1555+3653 & 25.49 &--23.3 & 2.7  & 1.63 &       &      &     &       &     & 1.12 & 0\\
J1604+3731 & 26.67 & ?  &3.7--3.9 & 1.50 & 0.289 & 7.73 & 3.7 & 0.068 & 1.5 & 1.05 & 5\\
J1615+3826 & 24.74 &--23.3 & 1.9  & 0.98 &       &      &     &       & 1.1 & .....& ...\\
J1635+3608 & 25.09 &--23.3 & 2.3  & 1.18 & 0.090 & 0.75 & 4.9 & 0.040 & ....& .....& ...\\
\hline
J1018--1240& 24.88 &--22.2 & 2.5  & 1.10 &       &      &     &       & ....& .....& ...\\
J1328--0307& 24.83 &--22.0 & 2.5  & 1.75 &       &      &     &       & ....& 1.15 & 5\\
J1457--0613& 25.82 &--23.3 & 3.0  & 0.87 &       &      &     &       & ....& 1.30 & 0\\
\hline
\end{tabular*}
\end{table*}

The above parameters are not estimated for a few faint sample sources for which
the total luminosity could not be determined because of a practically unknown
radio spectrum. It is worth  emphasizing that the parameters in columns 6, 7,
8, and 9 are calculated under  assumptions identical to those adopted by
Ishwara-Chandra \& Saikia (1999) in their list of previously known giant sources. Thus, the above parameters are homogeneously determined for our 
giants and theirs, and can be used for statistical analyses.
 
\subsection{Error estimates}
 
As in Machalski \& Jamrozy (2000), we have estimated the errors in the determination
of $B_{\rm me}$, $u_{\rm me}$, $B_{\rm iC}/B_{\rm me}$, and
$B^{2}_{\rm me}/(B^{2}_{\rm iC}+B^{2}_{\rm me})$ adopting errors of 20 per cent
in the integral radio luminosity, and 50 per cent in the source volume. These
give identical fractional errors for the above values as follows:
$B_{\rm me}\propto (1^{+0.27}_{-0.21})$, $u_{\rm me}\propto (1^{+0.60}_{-0.38})$,
$B_{\rm iC}/B_{\rm me}\propto (1^{+0.21}_{-0.27})$, and
$B^{2}_{\rm me}/(B^{2}_{\rm iC}+B^{2}_{\rm me})\propto (1^{+0.50}_{-0.40})$.
 
\subsection{Redshift and size estimates for identified faint galaxies}
 
The absolute $M_{\rm R}$ magnitude of the galaxies with spectroscopic redshift
varies from $-22.0$ mag to $-25.0$ mag (with the exception of J1428+2918 whose
apparent $R$ magnitude may be overestimated). A distribution of the $M_{\rm R}$
values in Table 6 has a mean of $-23.65$ mag and a standard deviation of 0.95
mag. Assuming this mean for the remaining identified galaxies but not observed
spectroscopically (cf. Table 1), one can estimate their redshift, and hence
distance, size, etc. Such estimates of redshift, logarithm of 1.4-GHz
luminosity, logarithm of radio--optical luminosity ratio (log\,$r$), and linear
size $D$ for the identified faint galaxies are given in columns 2, 3, 4, and 5
of Table 7. An uncertainty of the above estimates is computed taking into account
the minimal and maximal $M_{\rm R}$ values as $-22.0$ mag and $-25.0$ mag. Columns
6, 7, and 8 of Table 7 give the measure of asymmetries of the sources as do
columns 10, 11, and 12 of Table 6.
 
The size estimates of the sources in Table 7 are well above 1 Mpc, strongly
suggesting that all these sources would be giants even if their host galaxy had
the lowest absolute magnitude.
 
\begin{table*}
\caption{Probable physical parameters of the giant candidates with
photometric redshift estimates derived with the assumption of
$M_{\rm R}=(-23.65^{+1.65}_{-1.35})$ mag}
\begin{tabular*}{117mm}{lccccccc}
\hline
Source     &$(z\pm\Delta z)_{\rm est}$ & log$L_{\rm 1.4}$ & log $r$ &
$D$   & $S_{1}/S_{2}$ & $d_{2}/d_{1}$ & $\Delta$ \\
           &  & [W/Hz] &  & [Mpc] & & & [\degr]\\
\hline
J0927+3510 & $0.45^{+0.13}_{-0.17}$ & $26.0^{+0.3}_{-0.4}$ & 3.1 &
$2.3^{+0.3}_{-0.5}$ &     &(1.90)&(4)\\
           & $0.55^{+0.12}_{-0.15}$ & $26.2^{+0.2}_{-0.3}$ & 3.2 &
$2.5^{+0.2}_{-0.3}$ &     &(0.99)&(1.5)\\
J1011+3111 & $0.50^{+0.13}_{-0.15}$ & $25.9^{+0.2}_{-0.3}$ & 3.0 &
$2.0^{+0.2}_{-0.3}$ & 1.2 & 1.04 & 5\\
J1155+4029 & $0.53^{+0.12}_{-0.15}$ & $26.7^{+0.2}_{-0.3}$ & 3.7 &
$1.7^{+0.1}_{-0.2}$ & 6.5 & 1.95 & 2\\
J1200+3449 & $0.50^{+0.13}_{-0.15}$ & $26.5^{+0.2}_{-0.4}$ & 3.6 &
$1.0^{+0.1}_{-0.1}$ & 1.9 & 1.32 & 7\\
J1254+2933 & $0.42^{+0.12}_{-0.14}$ & $25.7^{+0.2}_{-0.4}$ & 2.7 &
$2.0^{+0.2}_{-0.4}$ & 1.9 & 1.24 & 1.5\\
J1355+2923 & $0.43^{+0.12}_{-0.14}$ & $26.2^{+0.2}_{-0.4}$ & 3.2 &
$1.7^{+0.2}_{-0.3}$ & 1.2 & 1.05 & 0\\
J1525+3345 & $0.47^{+0.12}_{-0.14}$ & $25.8^{+0.2}_{-0.3}$ & 2.8 &
$1.5^{+0.1}_{-0.2}$ & 2.3 & 1.03 & 3\\
J1554+3945 & $0.35^{+0.12}_{-0.13}$ & $25.6^{+0.3}_{-0.4}$ & 2.7 &
$1.3^{+0.2}_{-0.3}$ & 1.1 & 1.12 & 2\\
J1649+3114 & $0.36^{+0.12}_{-0.13}$ & $26.0^{+0.3}_{-0.4}$ & 3.0 &
$1.3^{+0.2}_{-0.3}$        &     & 1.61 & 1\\
J1712+3558 & $0.32^{+0.11}_{-0.12}$ & $25.6^{+0.3}_{-0.4}$ & 2.7 &
$1.2^{+0.2}_{-0.3}$        & 3.3 & 1.34 & 1\\
J1725+3923 & $0.29^{+0.11}_{-0.11}$ & $25.6^{+0.3}_{-0.4}$ & 2.6 &
$1.5^{+0.2}_{-0.4}$ & 0.8 & 1.36 & 1\\
\hline
\end{tabular*}
\end{table*}
 
\subsection{Remarks on some correlations between physical parameters}

\indent 
1) The investigated sample is inevitably biased against radio luminosity. The
sources at higher redshifts (or redshift estimates) are evidently more luminous.
This is not the case for the optical absolute magnitude; 11 galaxies with
$0.07<z<0.2$ in our sample have $\langle M_{\rm R}\rangle=(-23.5\pm 0.3)$ mag,
while 7 galaxies with $0.2<z<0.4$ have $\langle M_{\rm R}\rangle=-(23.7\pm 0.3)$
mag. The above parameters produce a spurious correlation between the 
radio--optical luminosity ratio (log\,$r$) and redshift.

2) The ratio $B_{\rm iC}/B_{\rm me}$ for all the giant sources in our sample
reaches the value of about 2 to 6 supporting the thesis that inverse-Compton
losses are a few times larger than synchrotron radiative losses in the time
evolution of the lobes of giant radio sources. Indeed, our limited data confirm
the correlations between the ratio $B_{\rm iC}/B_{\rm me}$ and linear size $D$,
as well as between $B^{2}_{\rm me}/(B^{2}_{\rm iC}+B^{2}_{\rm me})$ and $D$
pointed out by Ishwara-Chandra \& Saikia.
 
3) In most of the giant sources in Tables 6 and 7, the brighter lobe is closer 
to the host galaxy. This is expected if a source is intrinsically highly 
symmetric, its inclination angle from the observer's line of sight differs (but 
slightly) from $90\degr$, and it is at an  
age where the luminosity of its lobes is already decreasing. In such  cases, 
the asymmetries of the lobes in brightness and separation from the nucleus  
are generated by the different ages the lobes have in the observer's frame. The lobe 
(hotspots) seen as closer to the nucleus has to be younger and brighter than the 
farther one.

\begin{acknowledgements}
The authors acknowledge (i) the National Radio Astronomy Observatory (Socorro,
NM) for the target-of-opportunity observing time, (ii) the McDonald Observatory
(Mt. Locke, TX) for the observing time, (iii) the National Optical Astronomy
Observatories (Kitt Peak, AZ) for the usage of the IRAF software, (iv) the
Space Telescope Science Institute for the usage of the DSS data base,  and 
(v) Dr Luigina Feretti for her constructive remarks improving this paper. This work
was supported in part by the State Committee for Scientific Research
(KBN) under contract PB 0266/PO3/99/17.
\end{acknowledgements}

\clearpage 
\renewcommand{\thefigure}{1a}
\begin{figure*}[]
\resizebox{\hsize}{!}{\includegraphics{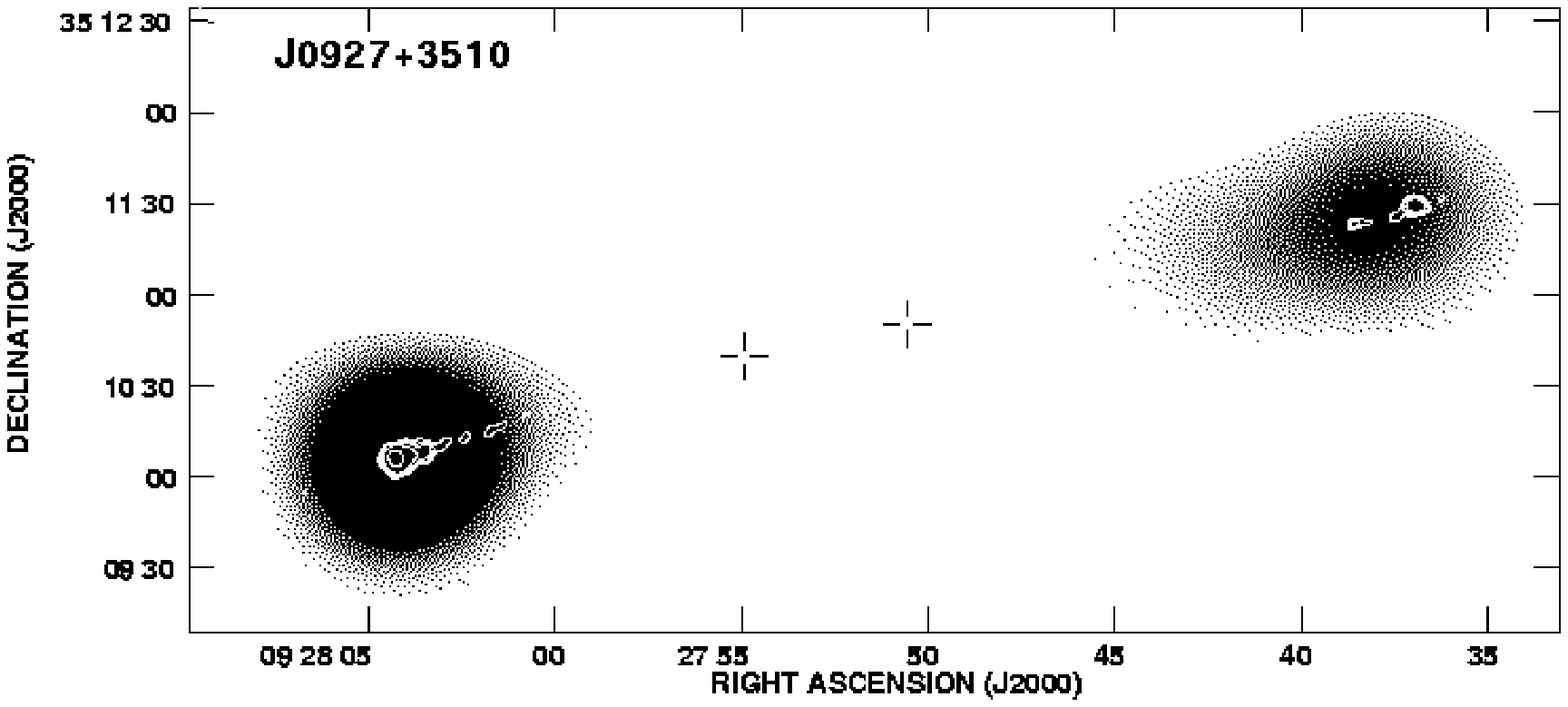}}
\caption{8.46-GHz VLA map of the source J0927+3510 with no radio core detected
(white contours) overlayed on 1.4-GHz NVSS map (gray scale). The crosses indicate
positions of the two galaxies given in Table~4 and shown in Fig.~3b}
\end{figure*}
 
\renewcommand{\thefigure}{1b}
\begin{figure*}
\resizebox{10cm}{!}{\includegraphics{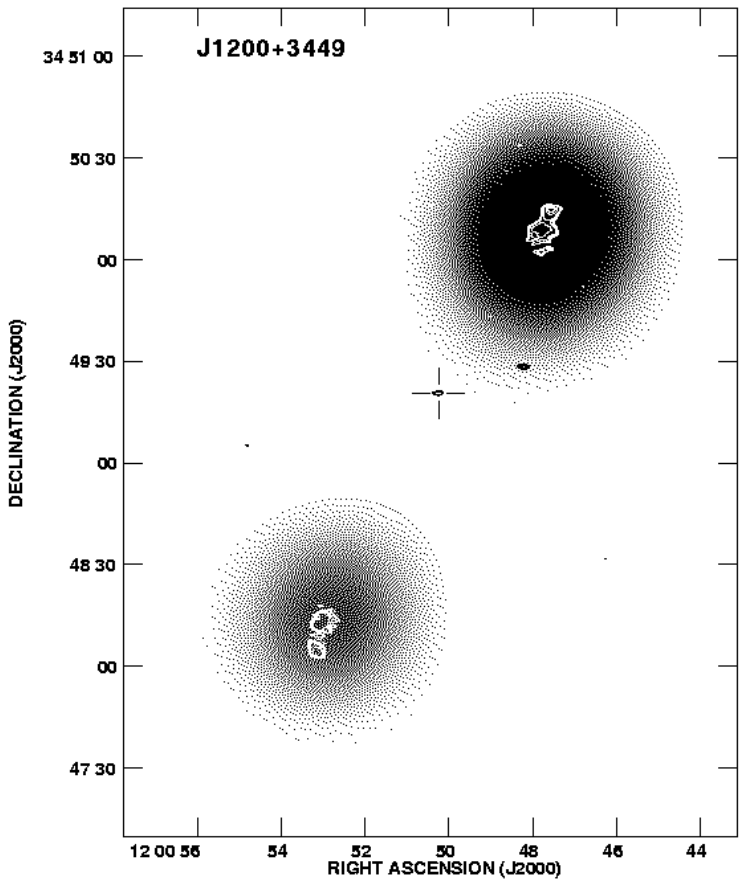}}
\hfill
\parbox[]{75mm}{
\caption{The same as in Fig.~1a but for the source J1200+3449. The radio core
detected is marked by black contours. The cross indicates position of the
identified galaxy (cf. Table~4)}}
\end{figure*}

\renewcommand{\thefigure}{1c}
\begin{figure*}
\resizebox{15cm}{!}{\includegraphics{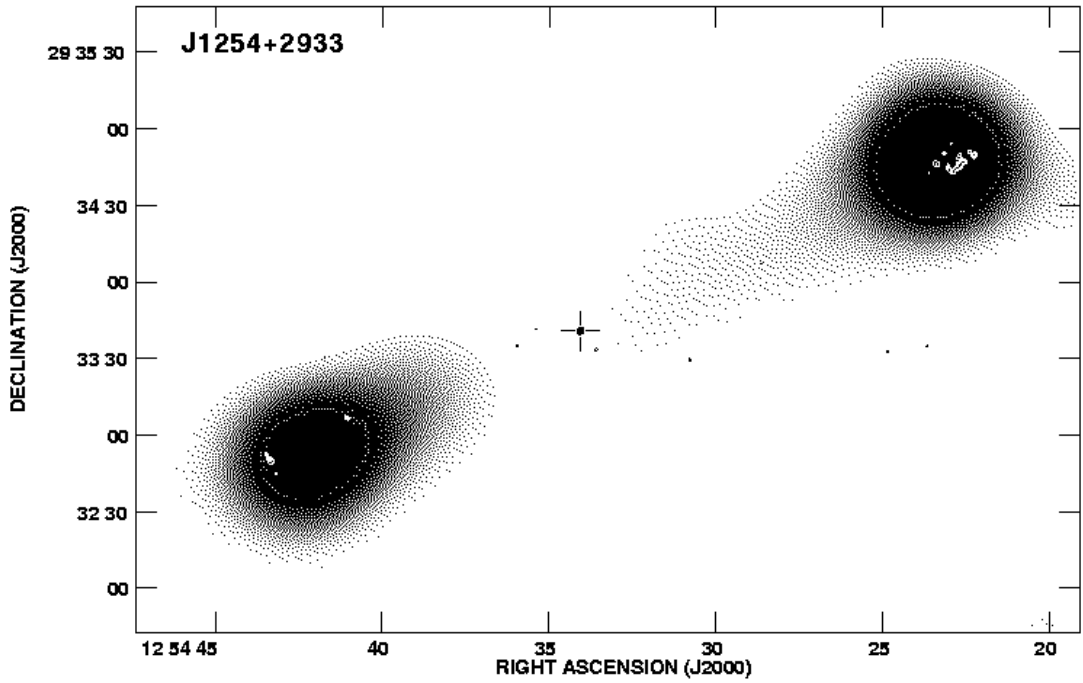}}
\caption{4.86-GHz VLA map of the source J1254+2933 (white contours) overlayed
on the NVSS map (gray scale). The cross indicates position of the radio core
and the identified galaxy}
\end{figure*}

\renewcommand{\thefigure}{1d}
\begin{figure*}
\resizebox{15cm}{!}{\includegraphics{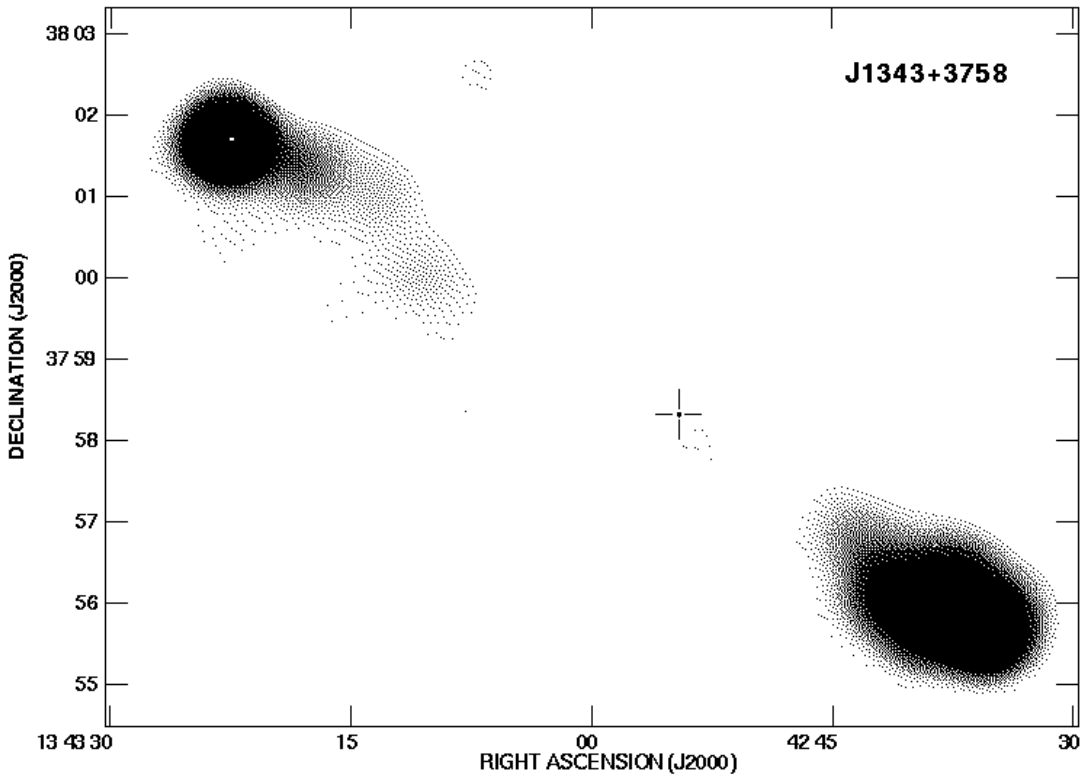}}
\caption{As in Fig.~1c but for the source J1343+3758}
\end{figure*}

\renewcommand{\thefigure}{1e}
\begin{figure*}
\resizebox{9cm}{!}{\includegraphics{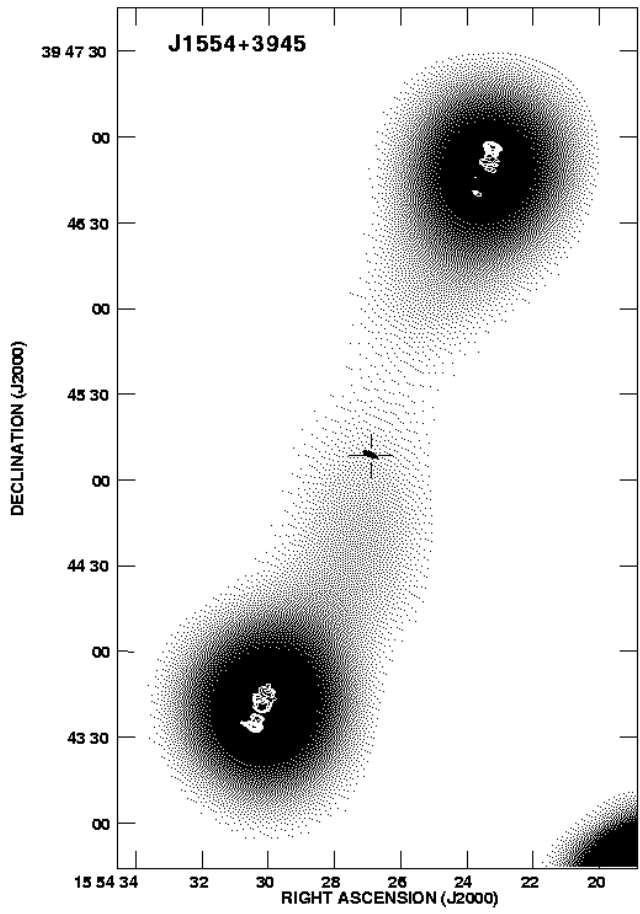}}
\hfill
\parbox[b]{85mm}{
\caption{As in Fig.~1c but for the source J1554+3945}}
\end{figure*}
 
\renewcommand{\thefigure}{1f}
\begin{figure*}
\resizebox{12cm}{!}{\includegraphics{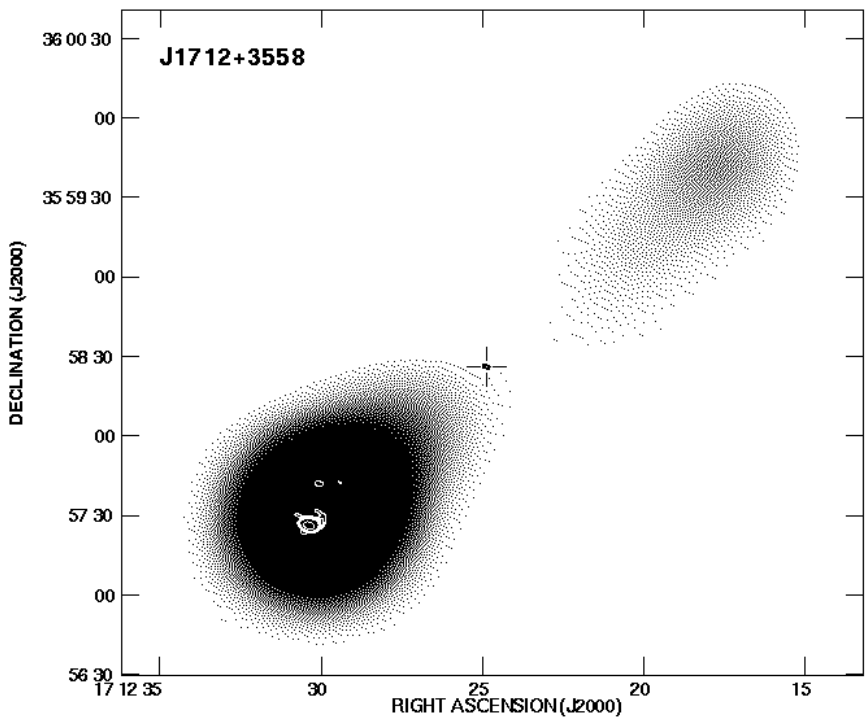}}
\hfill
\parbox[]{55mm}{
\caption{As in Fig.~1c but for the source J1712+3558}}
\end{figure*}

\renewcommand{\thefigure}{2a}
\begin{figure*}
\resizebox{11cm}{!}{\includegraphics{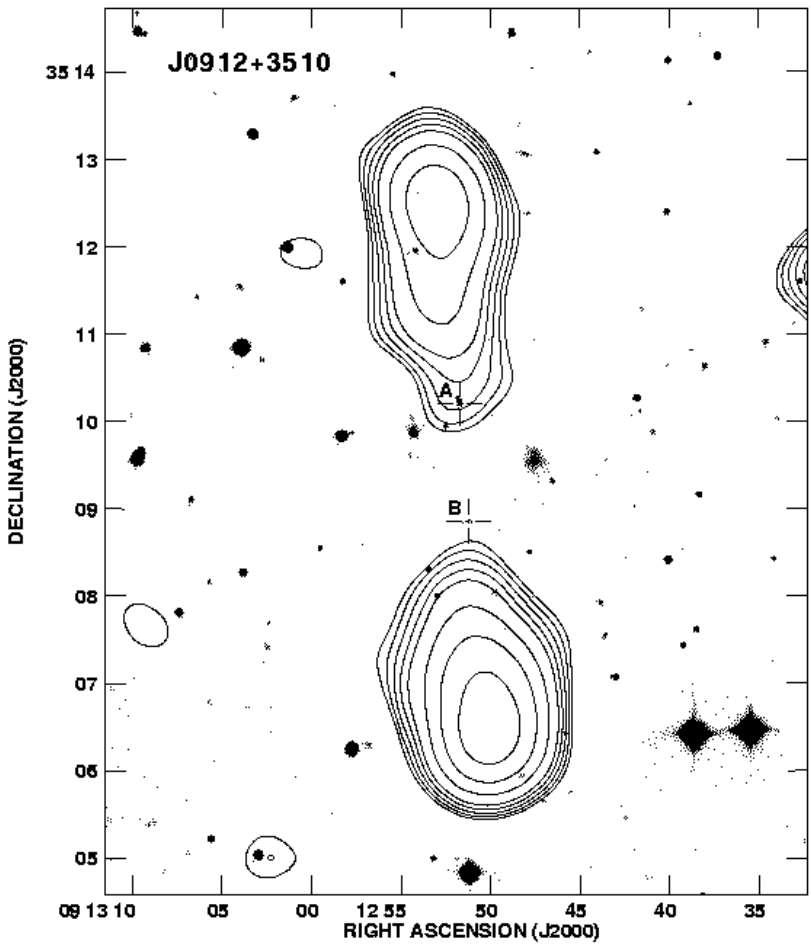}}
\hfill
\parbox[]{65mm}{
\caption{NVSS contour map of the source 0912+3510 overlayed on the optical
DSS image indicating the galaxies given in Table~4}}
\end{figure*}
 
\renewcommand{\thefigure}{2b}
\begin{figure*}
\resizebox{14cm}{!}{\includegraphics{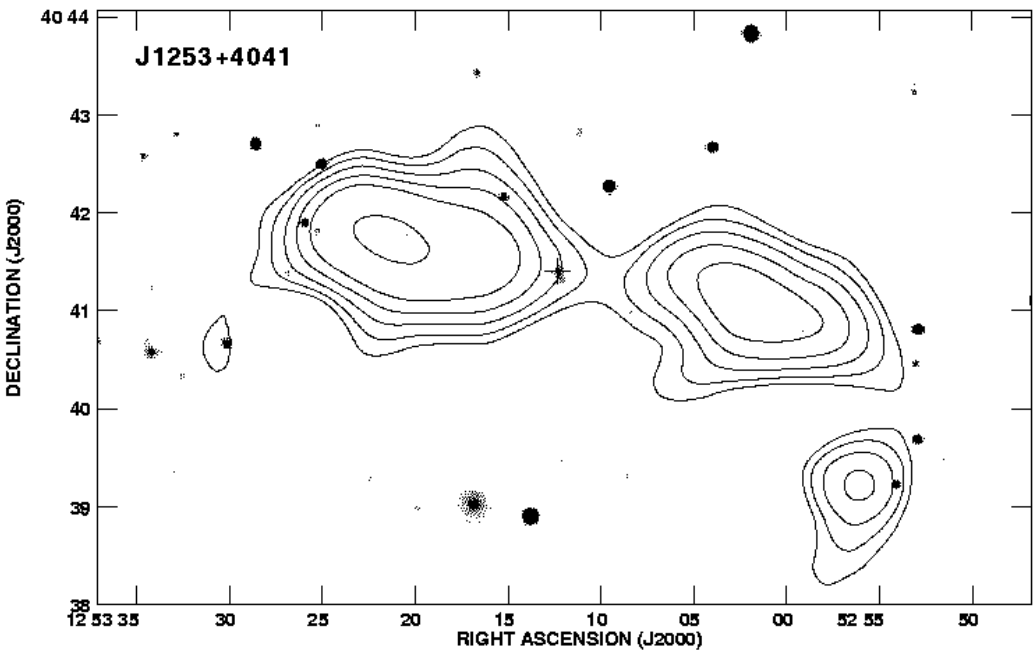}}
\caption{NVSS contour map of the source J1253+4041 overlayed on the optical
DSS image indicating the galaxy (cross) whose spectrum in shown in Fig.~4e}
\end{figure*}
 
\renewcommand{\thefigure}{2c}
\begin{figure*}
\resizebox{15cm}{!}{\includegraphics{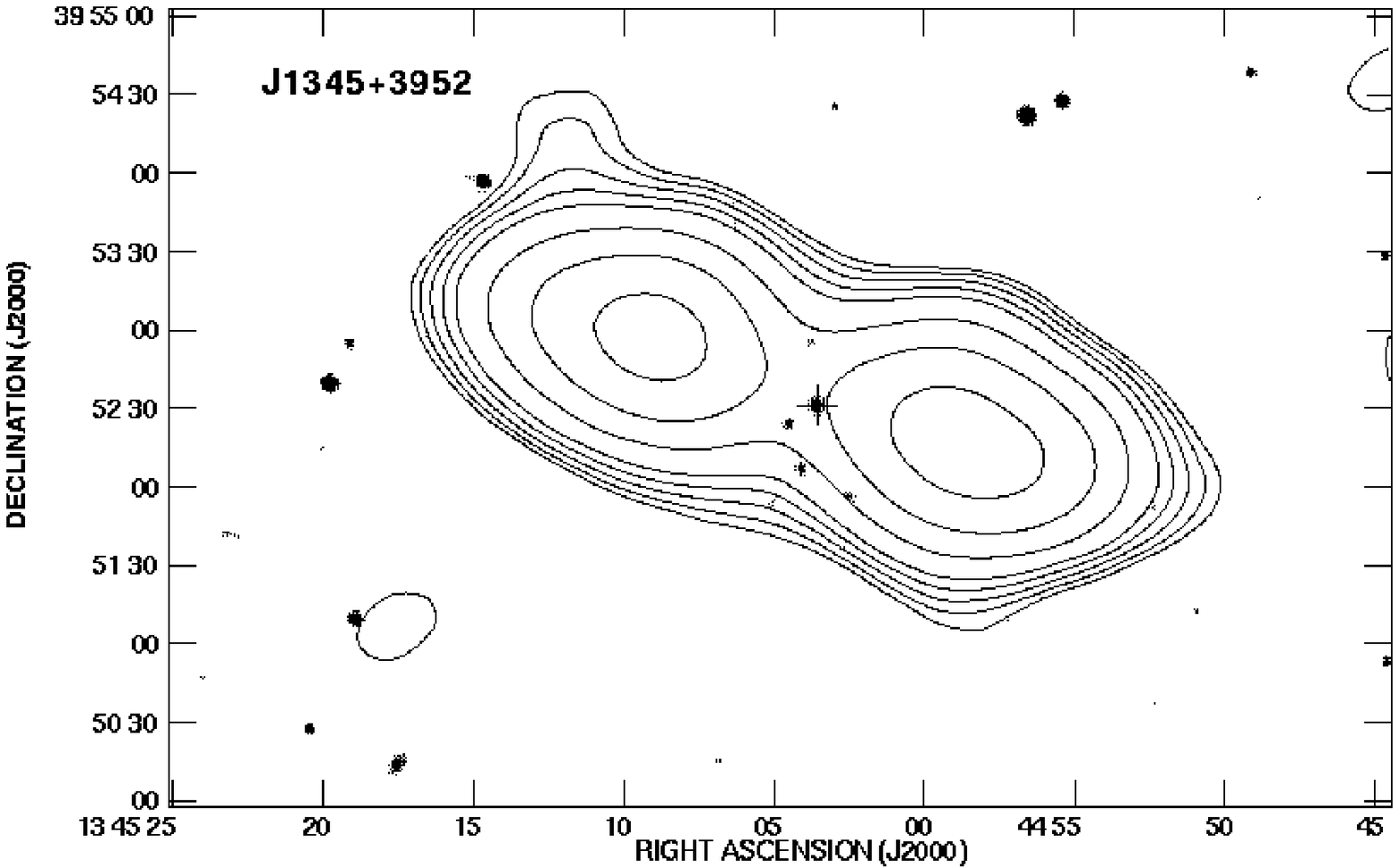}}
\caption{As in Fig.~2b but for the source J1345+3952. The spectrum of the
galaxy marked with the cross is shown in Fig.~4i}
\end{figure*}
 
\renewcommand{\thefigure}{2d}
\begin{figure*}
\resizebox{10cm}{!}{\includegraphics{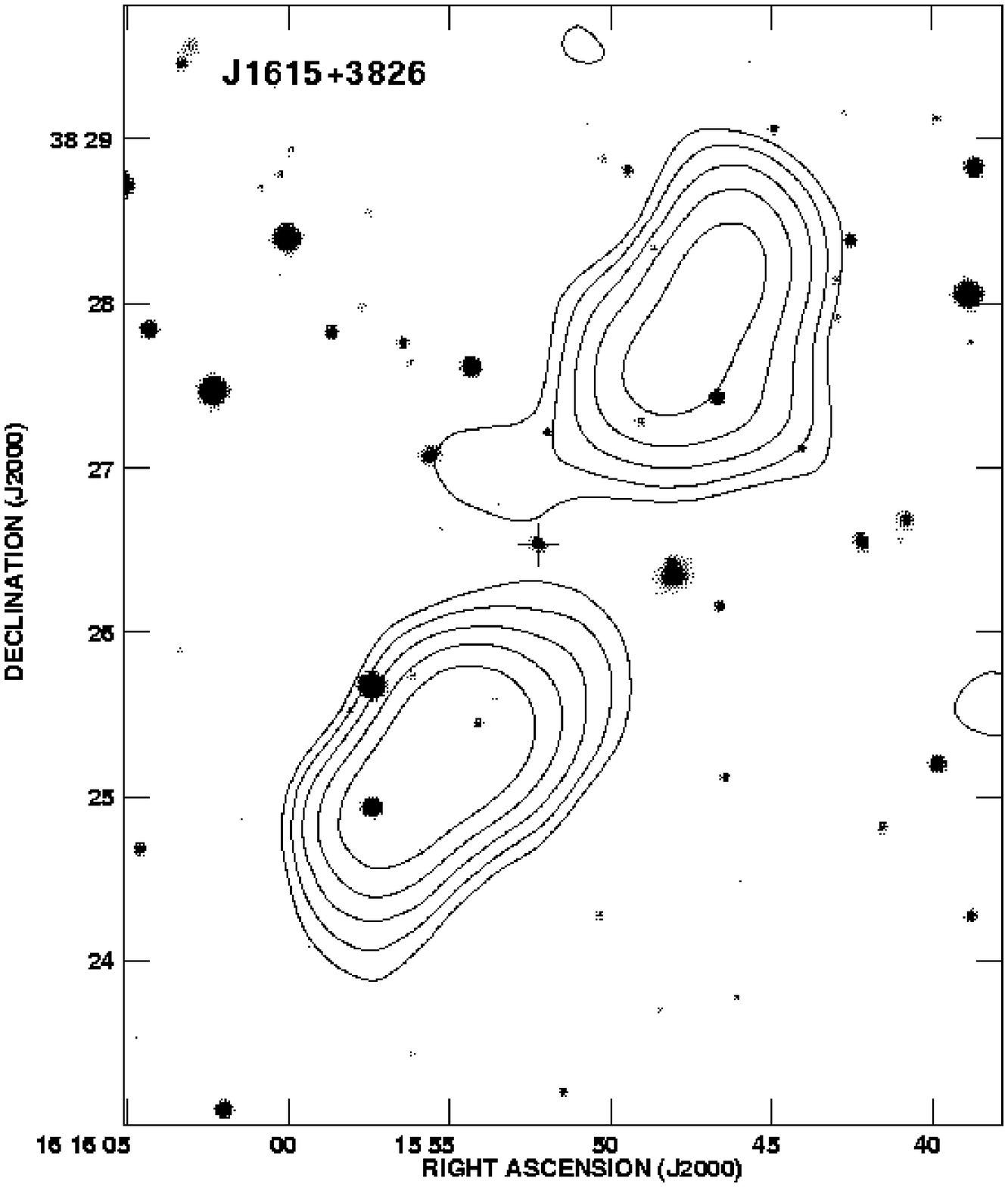}}
\hfill
\parbox[]{75mm}{
\caption{As in Fig.~2b but for the source J1615+3826. The spectrum of the
galaxy marked with the cross is shown in Fig.~4n}}
\end{figure*}

\renewcommand{\thefigure}{3a}
\begin{figure*}
\resizebox{12cm}{!}{\includegraphics{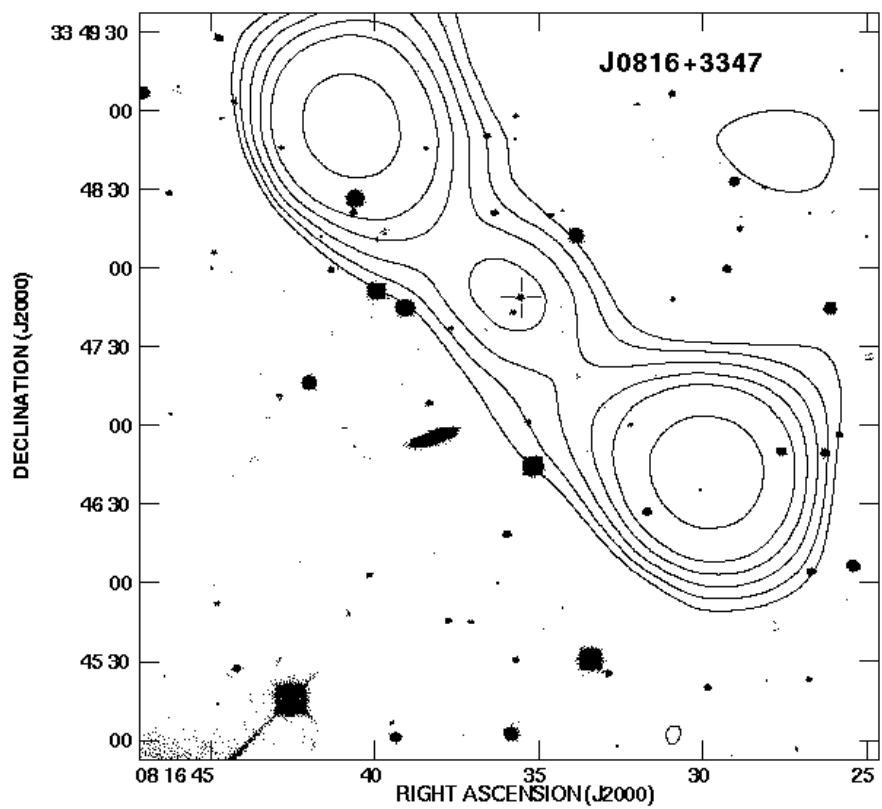}}
\hfill
\parbox[]{55mm}{
\caption{Deep $R$-band image of the optical field around the source J0816+3347.
The host optical object (possibly a quasar) is marked with the cross. The NVSS
contour map is overlayed for a comparison}}
\end{figure*}
 
\renewcommand{\thefigure}{3b}
\begin{figure*}
\resizebox{12cm}{!}{\includegraphics{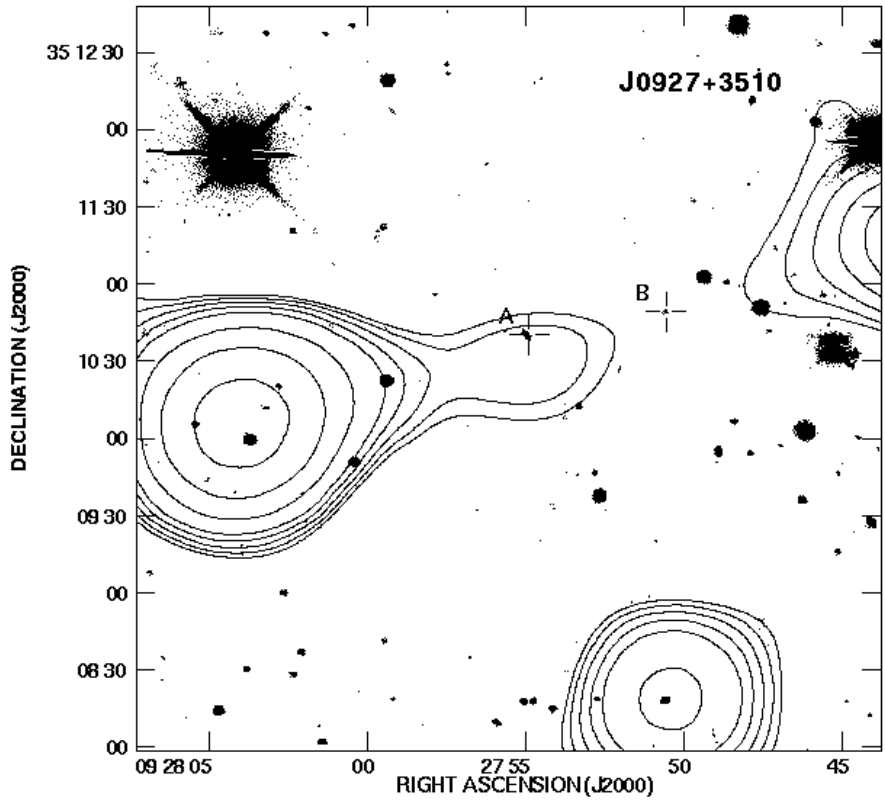}}
\hfill
\parbox[]{55mm}{
\caption{As in Fig.~3a but around the source J0927+3510. The two galaxies listed
in Table~4 are marked with the crosses}}
\end{figure*}
 
\renewcommand{\thefigure}{3c}
\begin{figure*}
\resizebox{12cm}{!}{\includegraphics{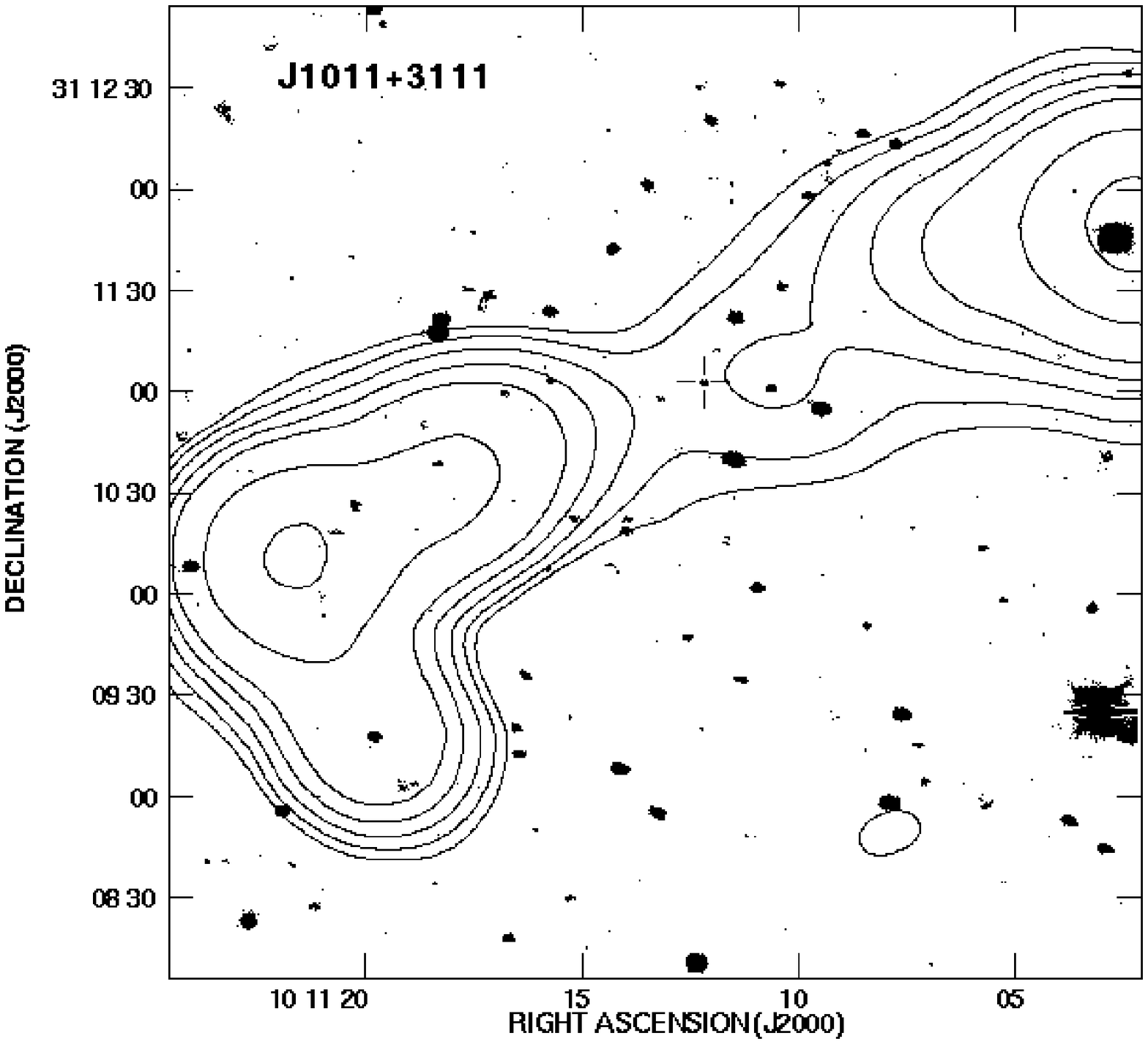}}
\hfill
\parbox[]{55mm}{
\caption{As in Fig.~3a but around the source J1011+3111. The identified 21.2 mag
galaxy in marked with the cross}}
\end{figure*}
 
\renewcommand{\thefigure}{3d}
\begin{figure*}
\resizebox{12cm}{!}{\includegraphics{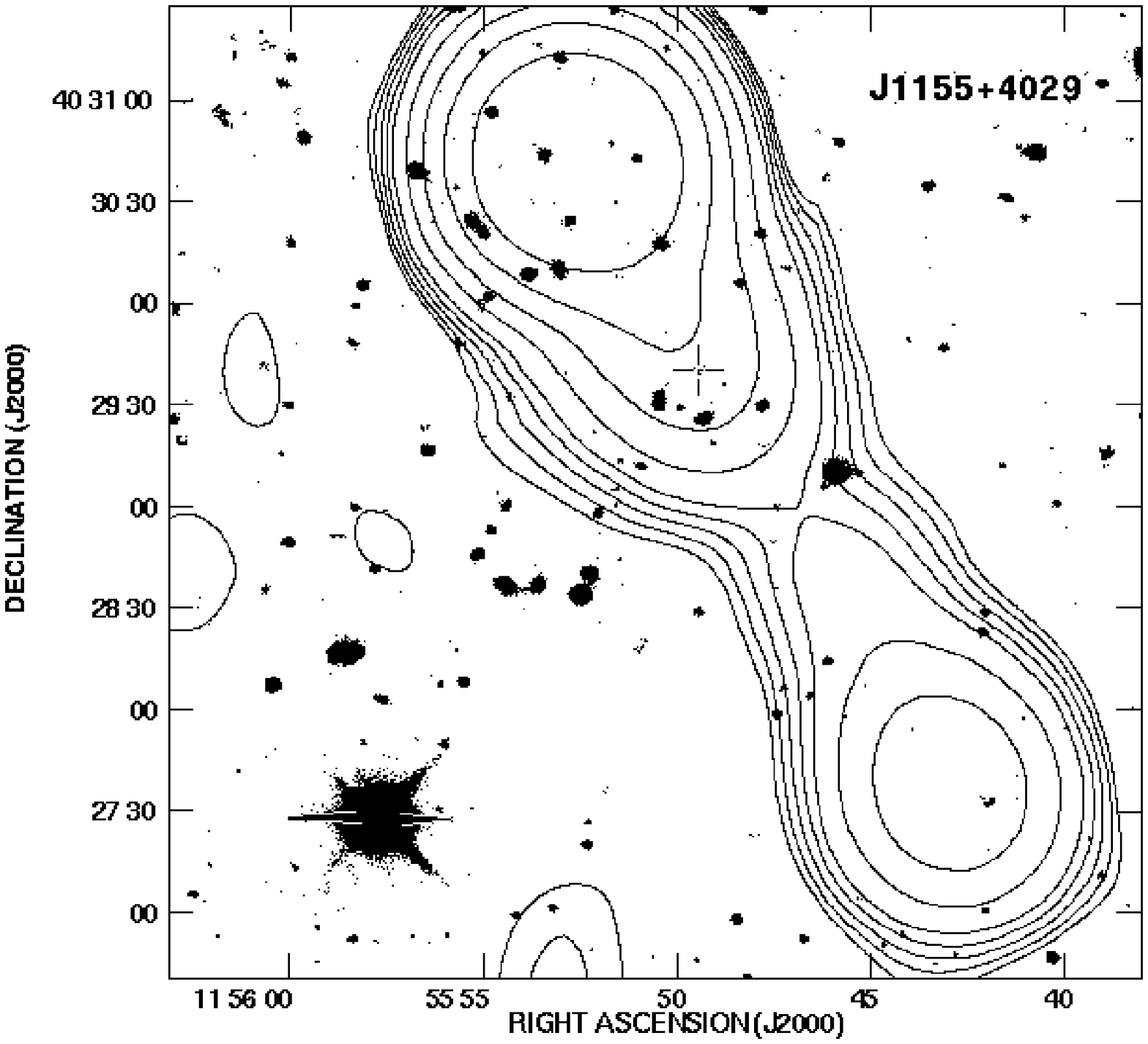}}
\hfill
\parbox[]{55mm}{
\caption{As in Fig.~3a but around the source J1155+4029. The identified 21.5 mag
galaxy is marked with the cross}}
\end{figure*}

\renewcommand{\thefigure}{3e}
\begin{figure*}
\resizebox{12cm}{!}{\includegraphics{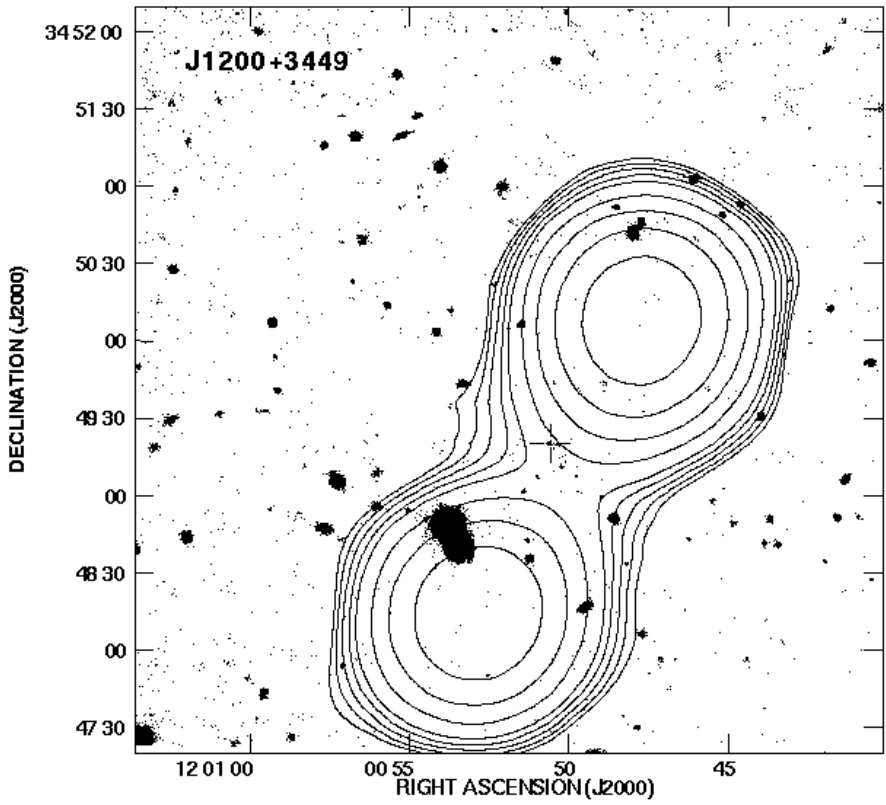}}
\hfill
\parbox[]{55mm}{
\caption{As in Fig.~3a but around the source J1200+3449. The identified 21.2 mag
galaxy  and the radio core (cf. Fig.~1b) are deflected from the axis of radio
cocoon suggested by the NVSS contour map overlayed}}
\end{figure*}

\renewcommand{\thefigure}{3f} 
\begin{figure*}
\resizebox{12cm}{!}{\includegraphics{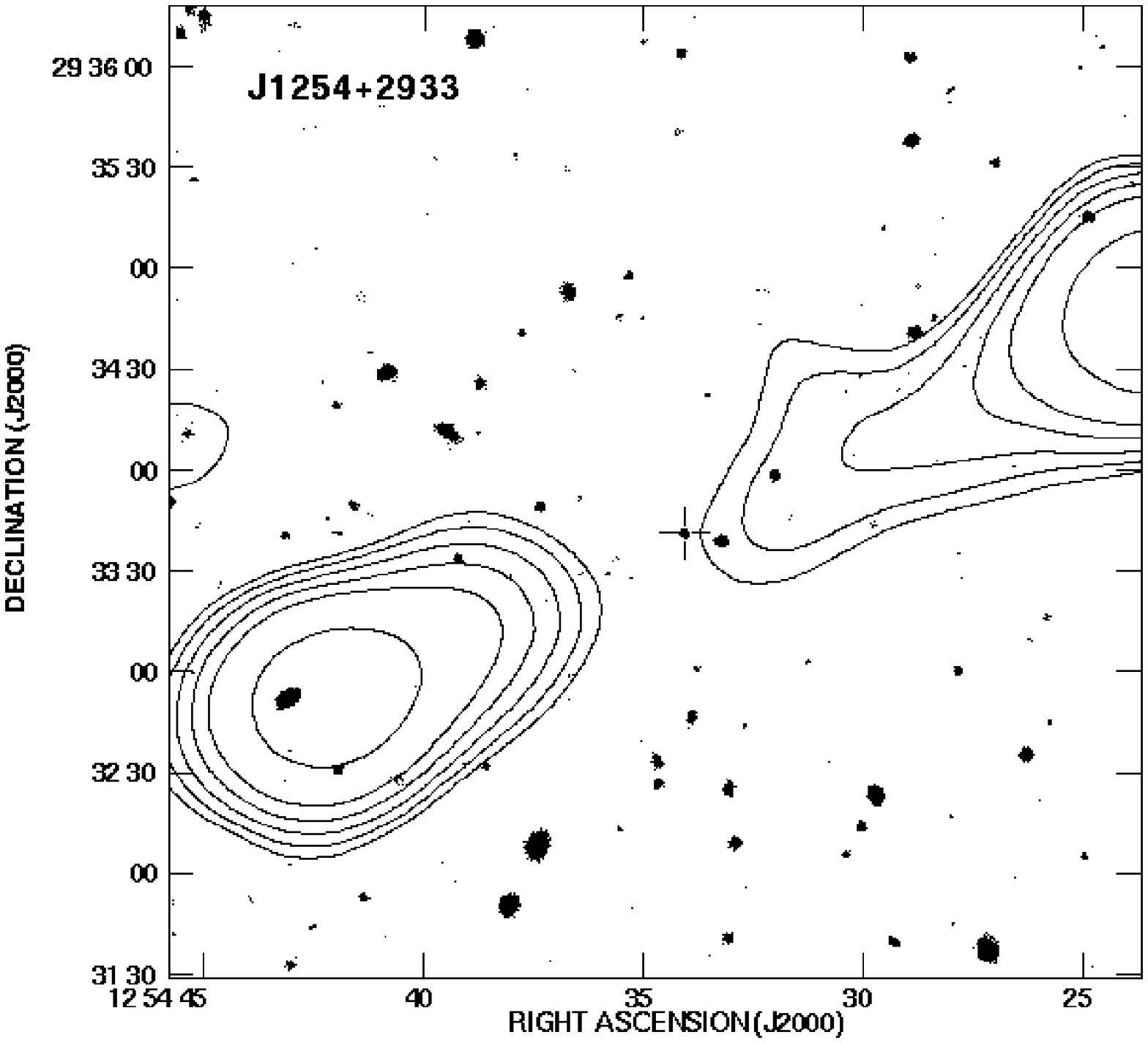}}
\hfill
\parbox[]{55mm}{
\caption{As in Fig.~3a but around the source J1254+2933. The identified 20.3 mag
galaxy is marked with the cross}}
\end{figure*}

\renewcommand{\thefigure}{3g} 
\begin{figure*}
\resizebox{12cm}{!}{\includegraphics{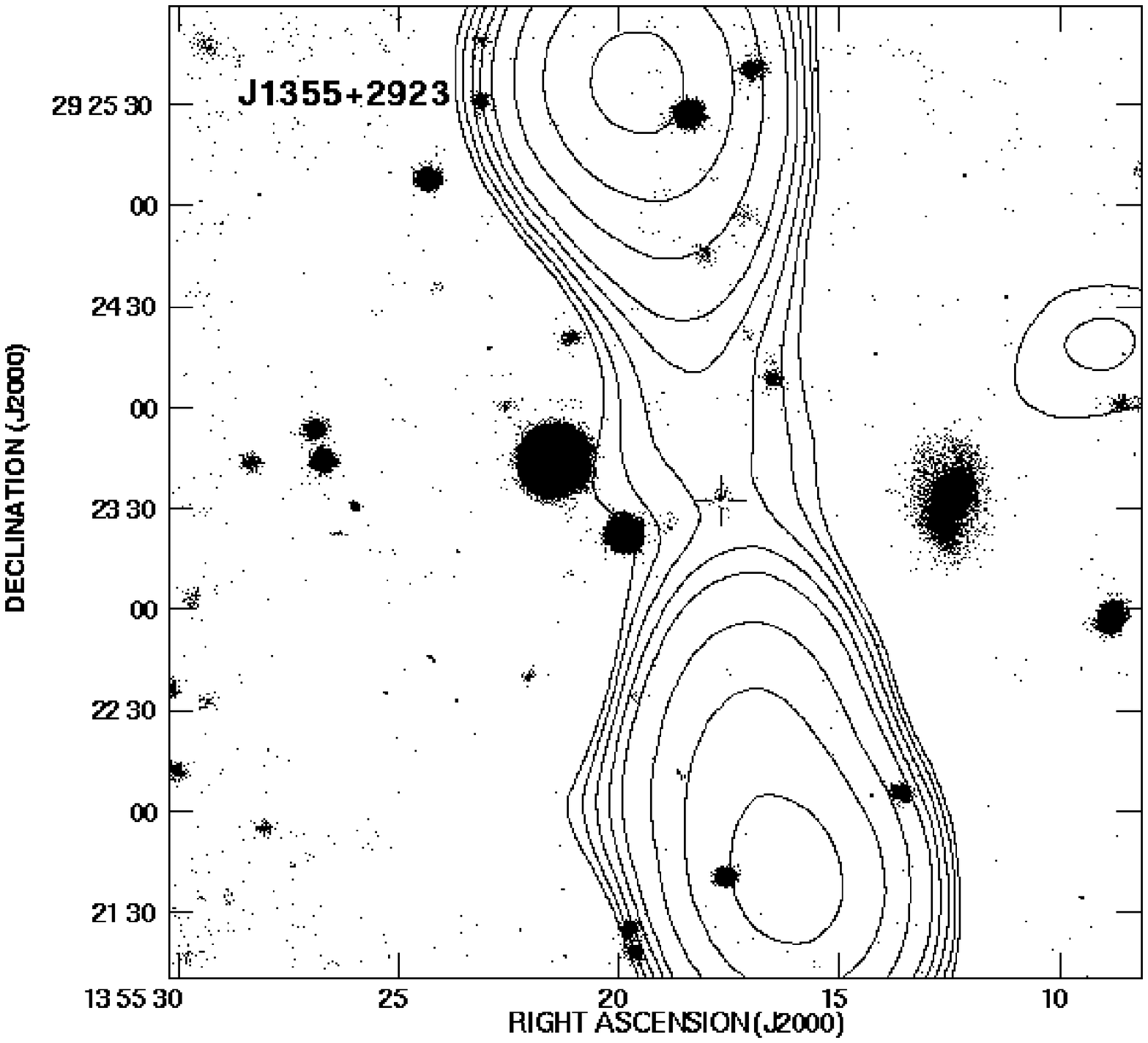}}
\hfill
\parbox[]{55mm}{
\caption{As in Fig.~3a but around the source J1355+2923. The identified 20.4 mag
galaxy is marked with the cross}}
\end{figure*}

\renewcommand{\thefigure}{3h} 
\begin{figure*}
\resizebox{12cm}{!}{\includegraphics{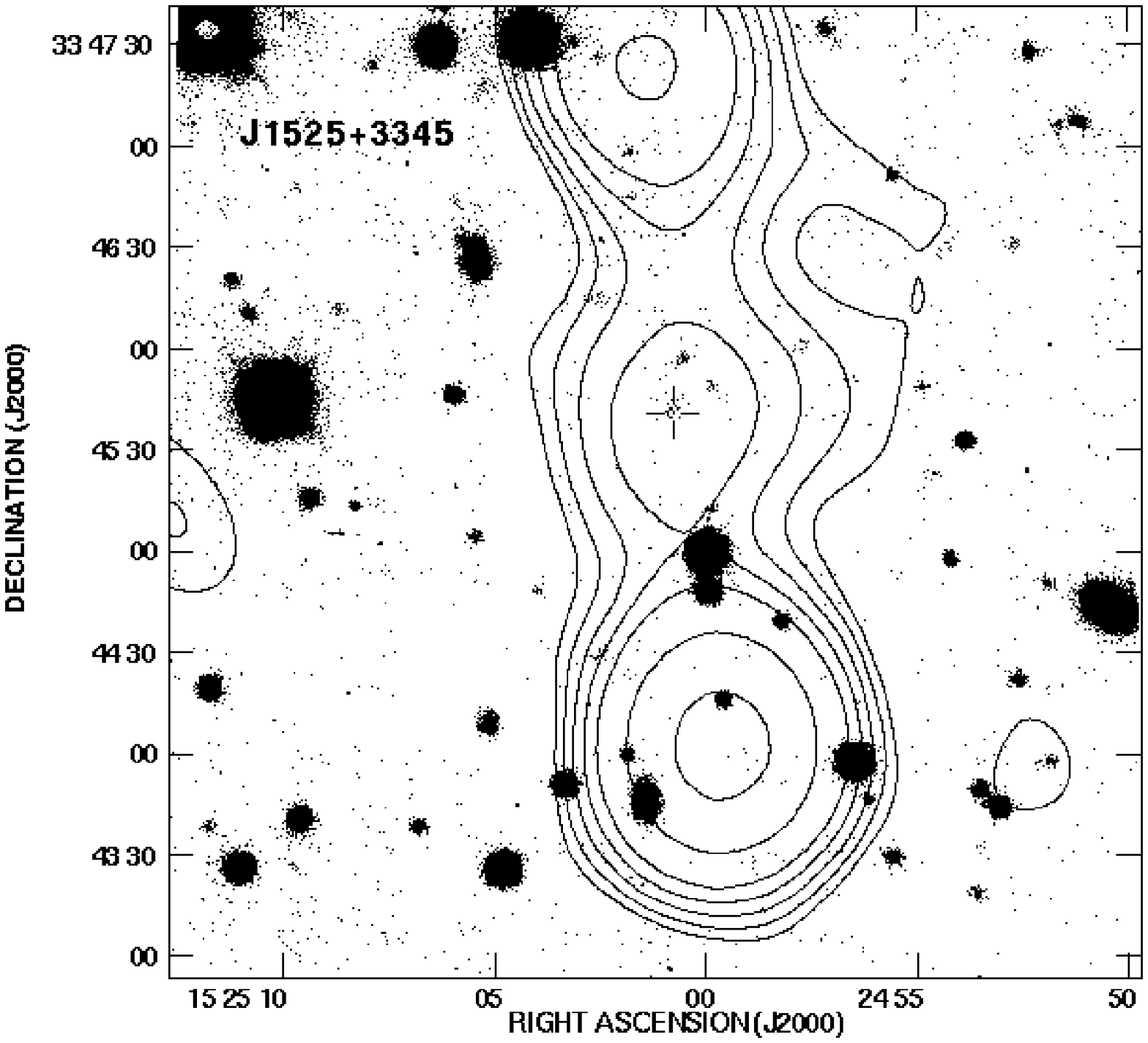}}
\hfill
\parbox[]{55mm}{
\caption{As in Fig.~3a but around the source J1525+3345. The identified 20.9 mag
galaxy is marked with the cross}}
\end{figure*}
 
\clearpage
\cleardoublepage

\renewcommand{\thefigure}{4a}
\begin{figure*}
\resizebox{15cm}{!}{\includegraphics{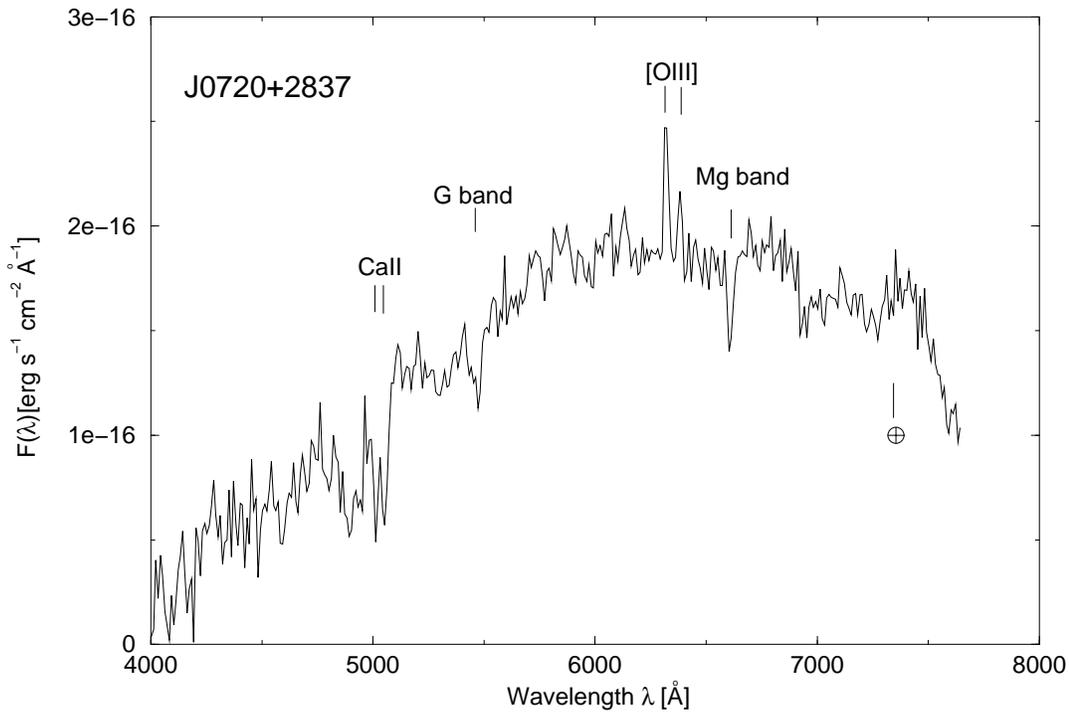}}
\caption{Optical spectrum of the galaxy identified with the source J0720+2837}
\end{figure*}

\renewcommand{\thefigure}{4b}
\begin{figure*}
\resizebox{15cm}{!}{\includegraphics{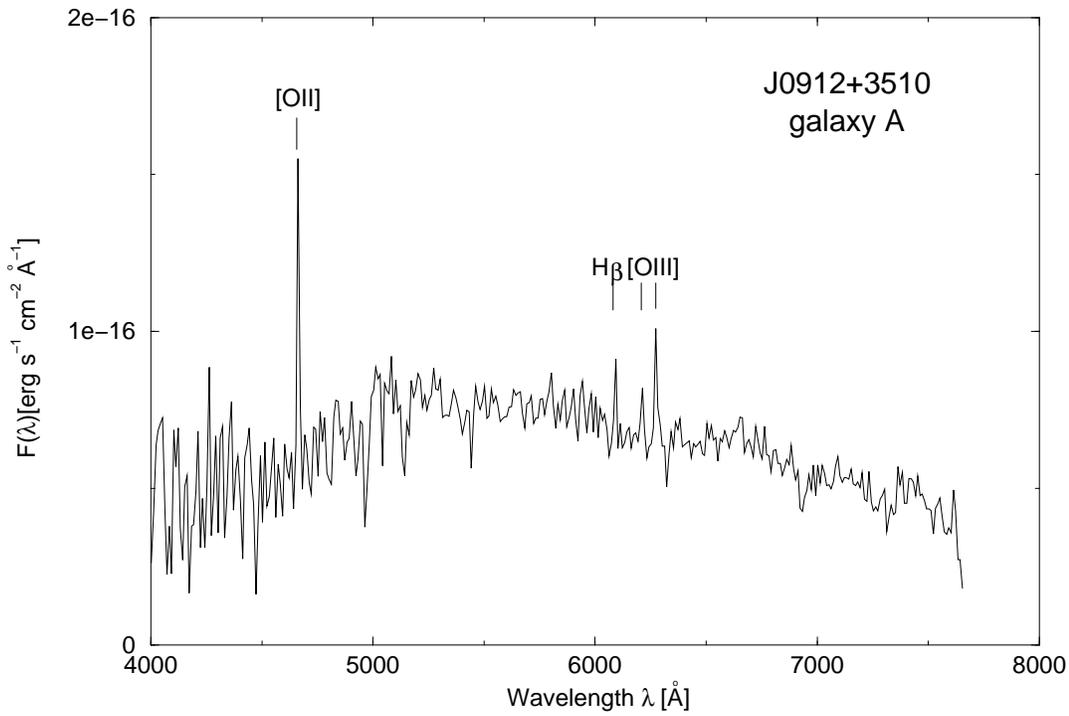}}
\caption{Optical spectrum of the galaxy A, a possible identification of the
source J0912+3510}
\end{figure*}

\renewcommand{\thefigure}{4c}
\begin{figure*}
\resizebox{15cm}{!}{\includegraphics{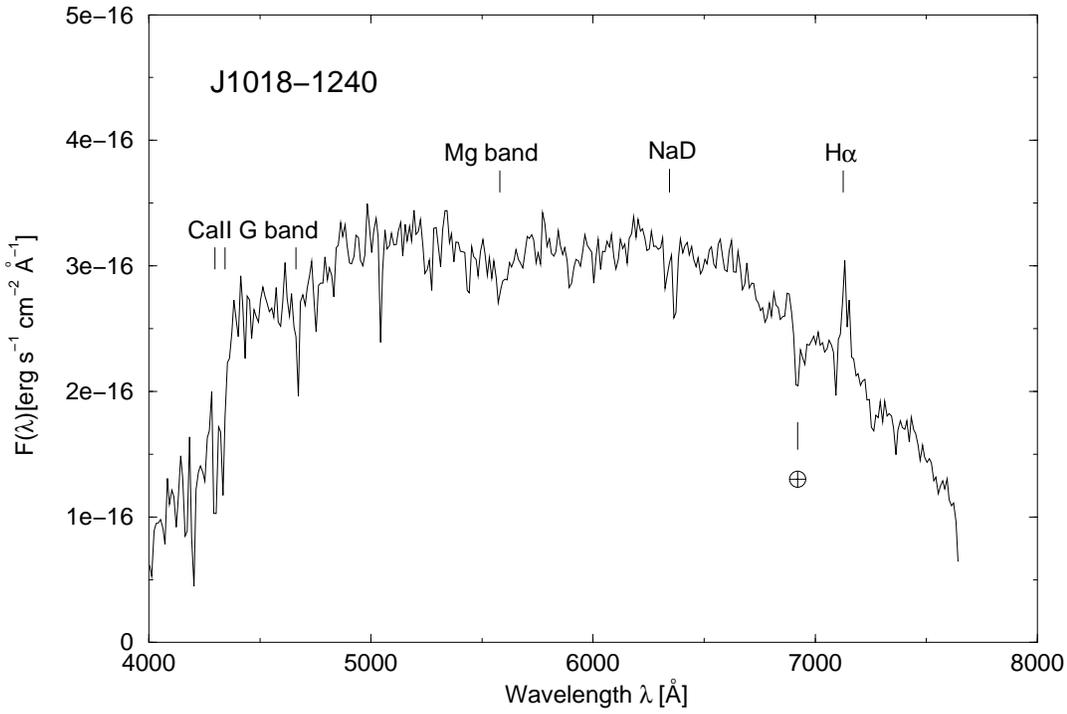}}
\caption{As in Fig.~4a but for the galaxy identified with the source J1018--1240}
\end{figure*}
 
\renewcommand{\thefigure}{4d}
\begin{figure*}
\resizebox{15cm}{!}{\includegraphics{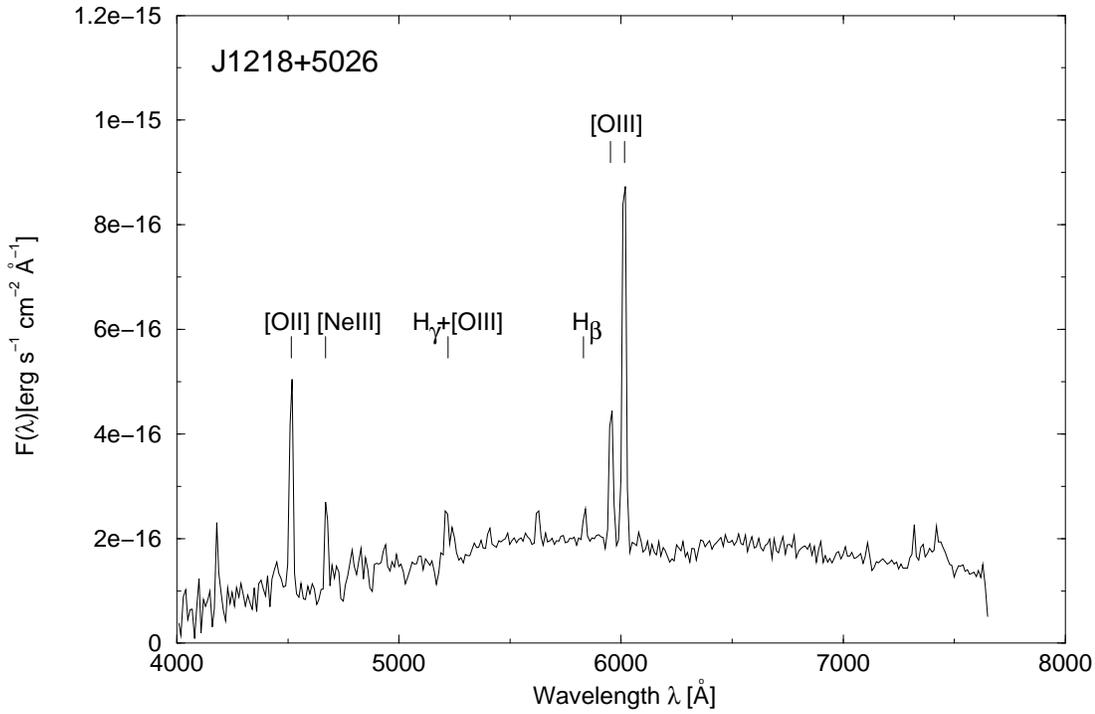}}
\caption{As in Fig.~4a but for the galaxy identified with the source J1218+5026}
\end{figure*}
 
\renewcommand{\thefigure}{4e}
\begin{figure*}
\resizebox{15cm}{!}{\includegraphics{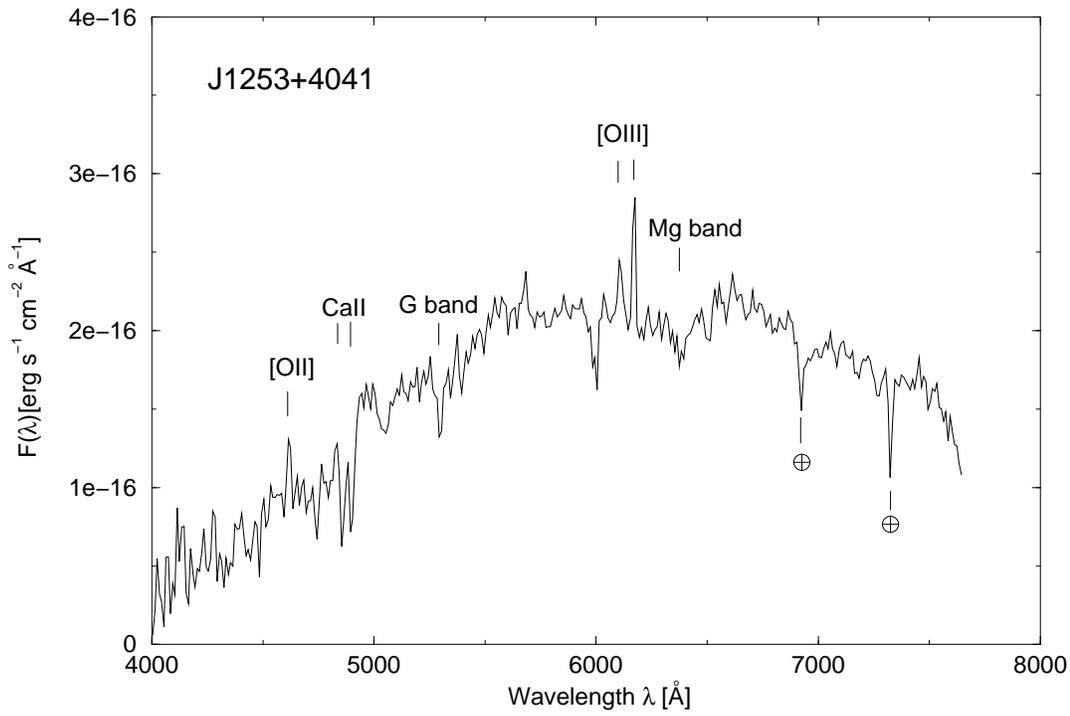}}
\caption{As in Fig.~4a but for the galaxy likely identified with the source
J1254+4041}
\end{figure*}
 
\renewcommand{\thefigure}{4f}
\begin{figure*}
\resizebox{15cm}{!}{\includegraphics{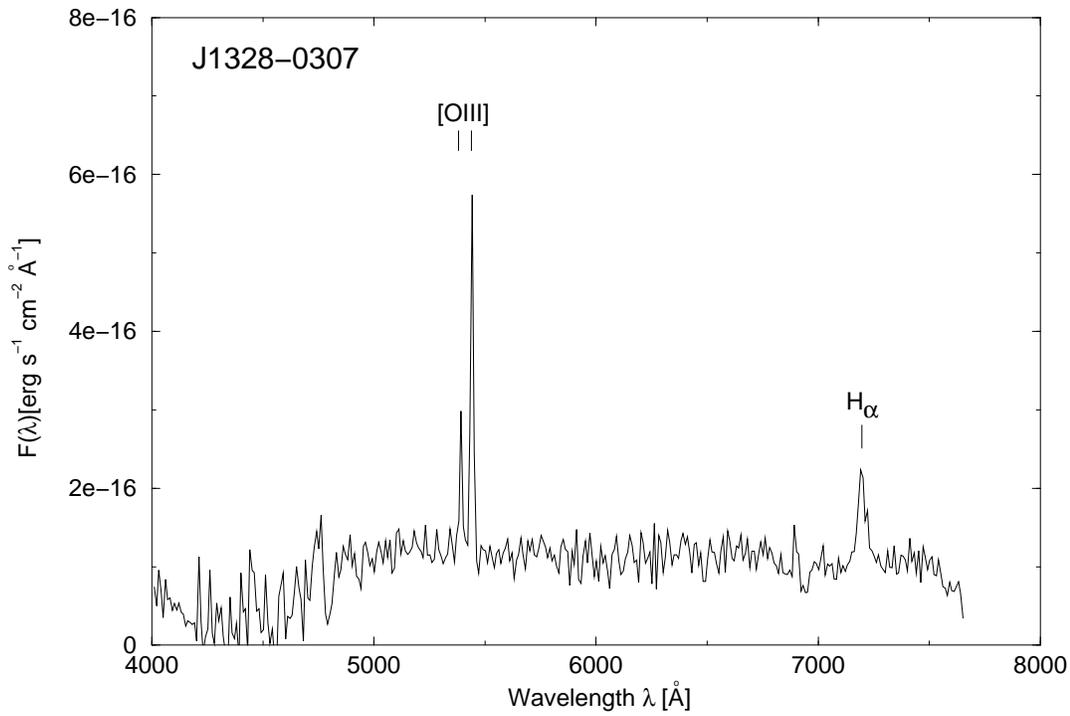}}
\caption{As in Fig.~4a but for the galaxy identified with the source J1328--0307}
\end{figure*}
 
\renewcommand{\thefigure}{4g}
\begin{figure*}
\resizebox{15cm}{!}{\includegraphics{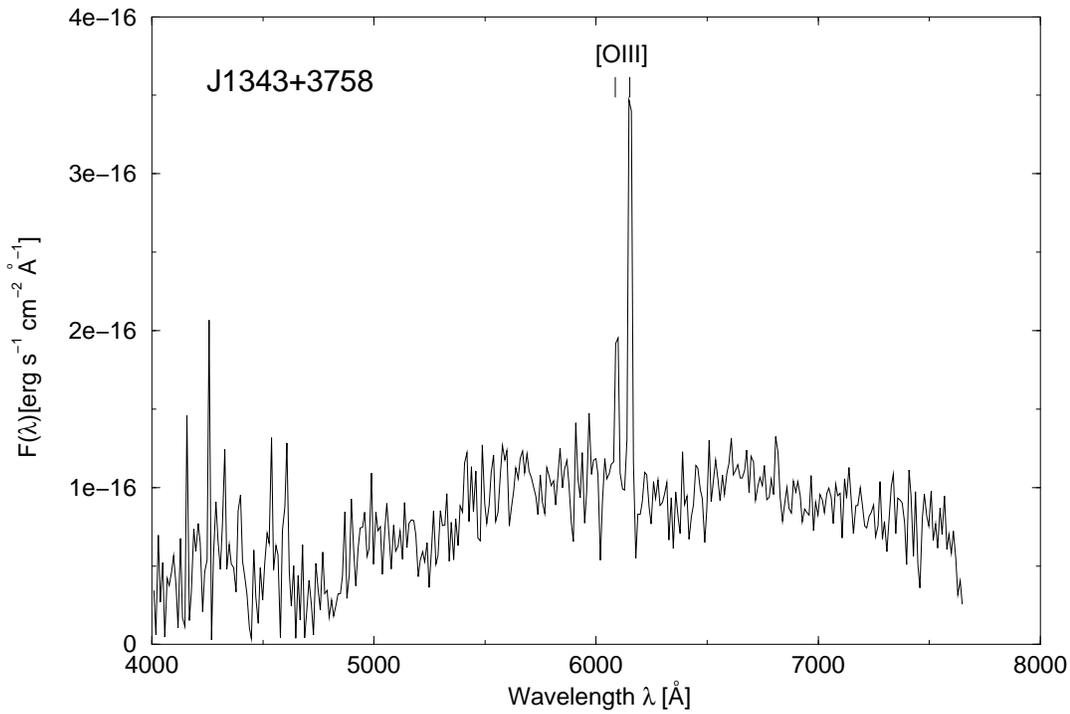}}
\caption{As in Fig.~4a but for the galaxy identified with the source J1343+3758}
\end{figure*}
 
\renewcommand{\thefigure}{4h}
\begin{figure*}
\resizebox{15cm}{!}{\includegraphics{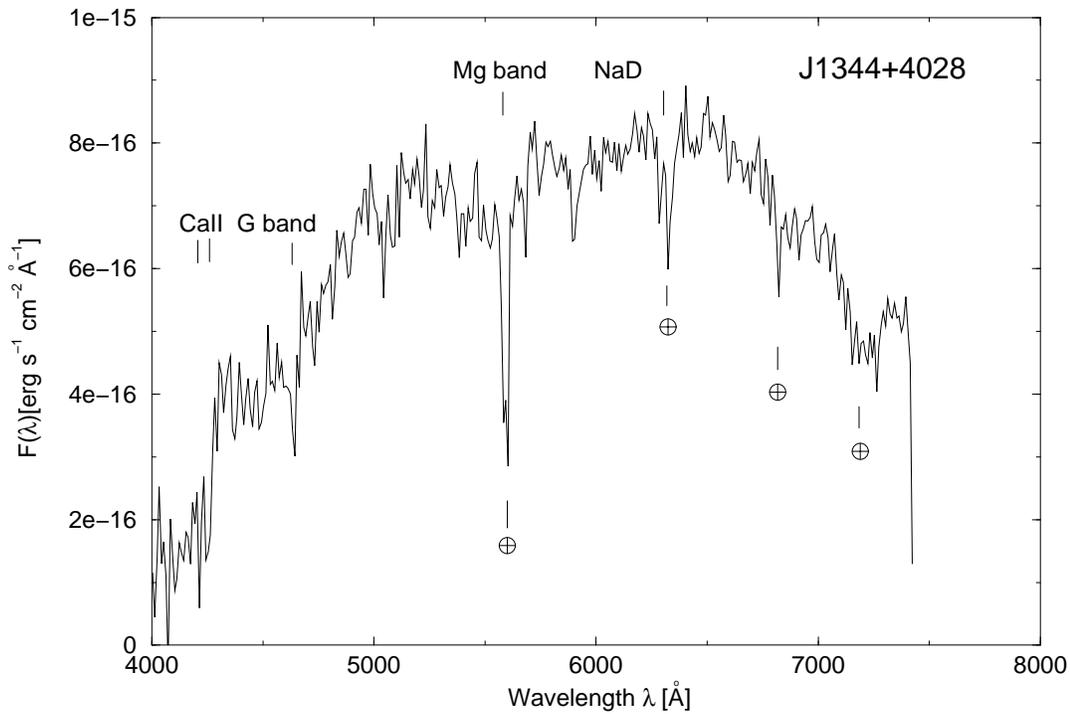}}
\caption{As in Fig.4a but for the galaxy identified with the source J1344+4028}
\end{figure*}
 
\renewcommand{\thefigure}{4i}
\begin{figure*}
\resizebox{15cm}{!}{\includegraphics{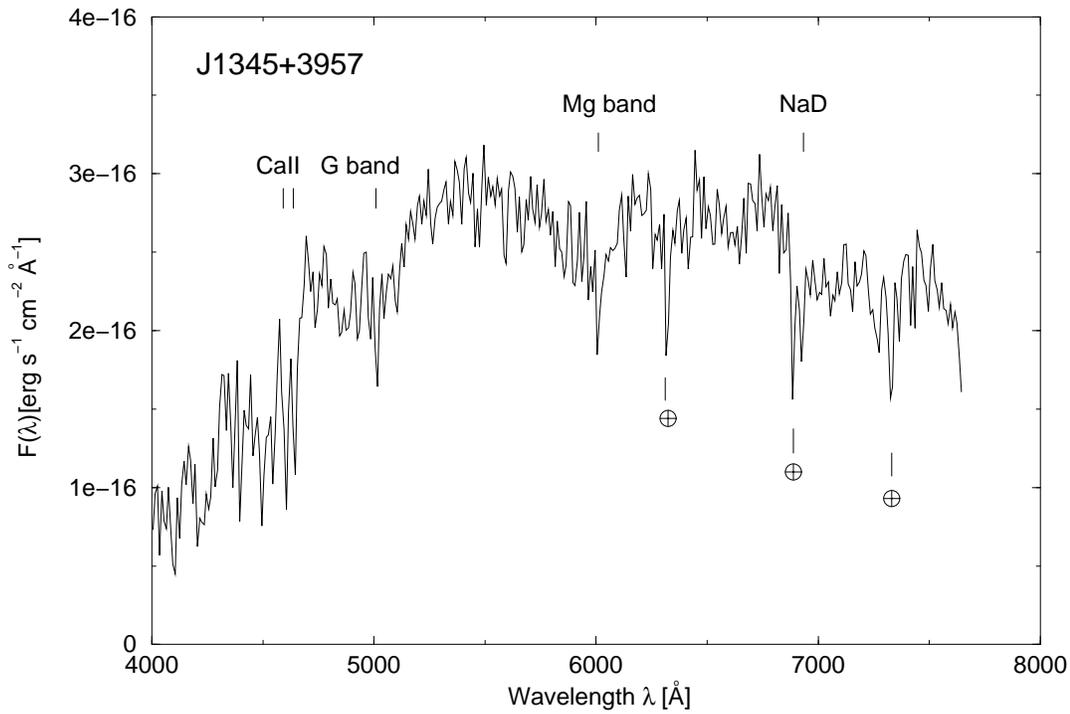}}
\caption{As in Fig.~4a but for the galaxy likely identified with the source
J1345+3952}
\end{figure*}
 
\renewcommand{\thefigure}{4j}
\begin{figure*}
\resizebox{15cm}{!}{\includegraphics{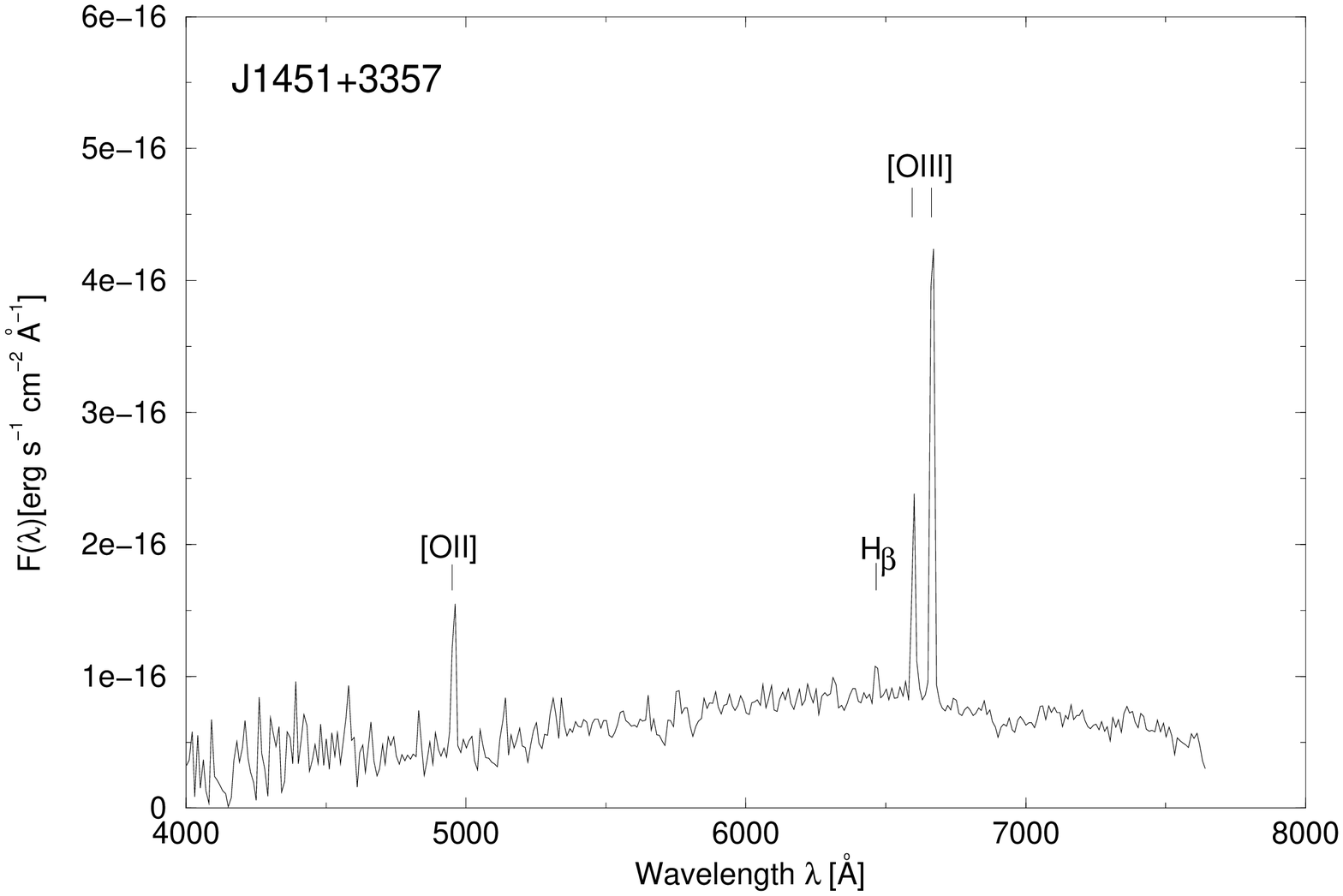}}
\caption{As in Fig.~4a but for the galaxy identified with the source J1451+3357}
\end{figure*}
 
\renewcommand{\thefigure}{4k}
\begin{figure*}
\resizebox{15cm}{!}{\includegraphics{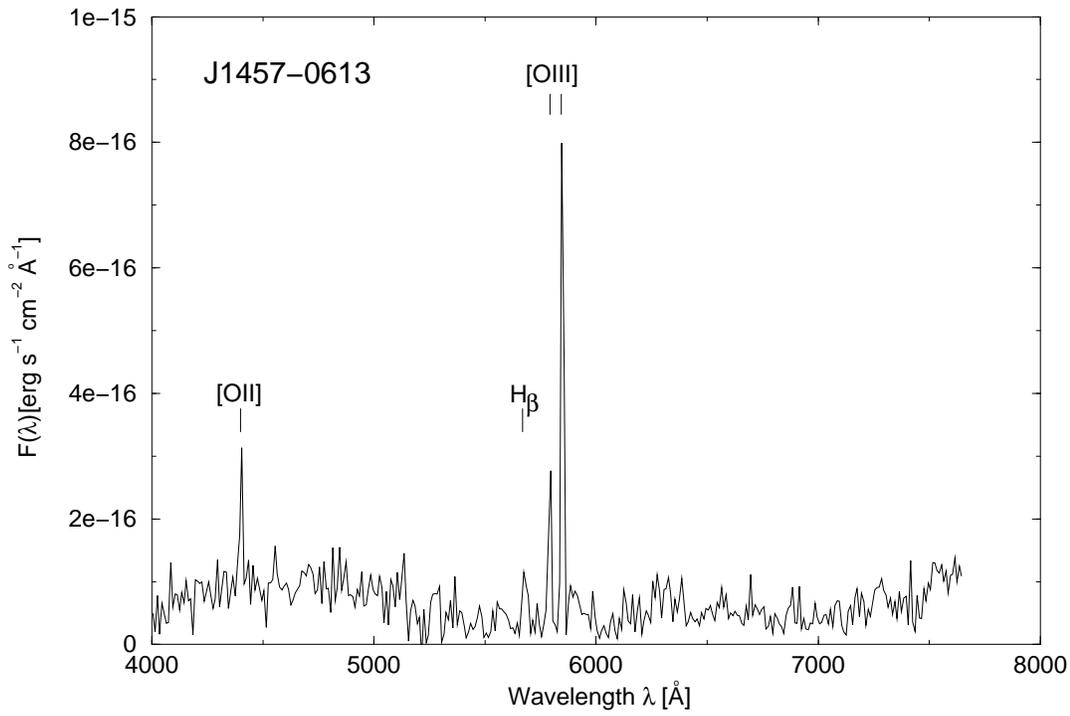}}
\caption{As in Fig.~4a but for the galaxy identified with the source J1457--0613}
\end{figure*}
 
\renewcommand{\thefigure}{4l}
\begin{figure*}
\resizebox{15cm}{!}{\includegraphics{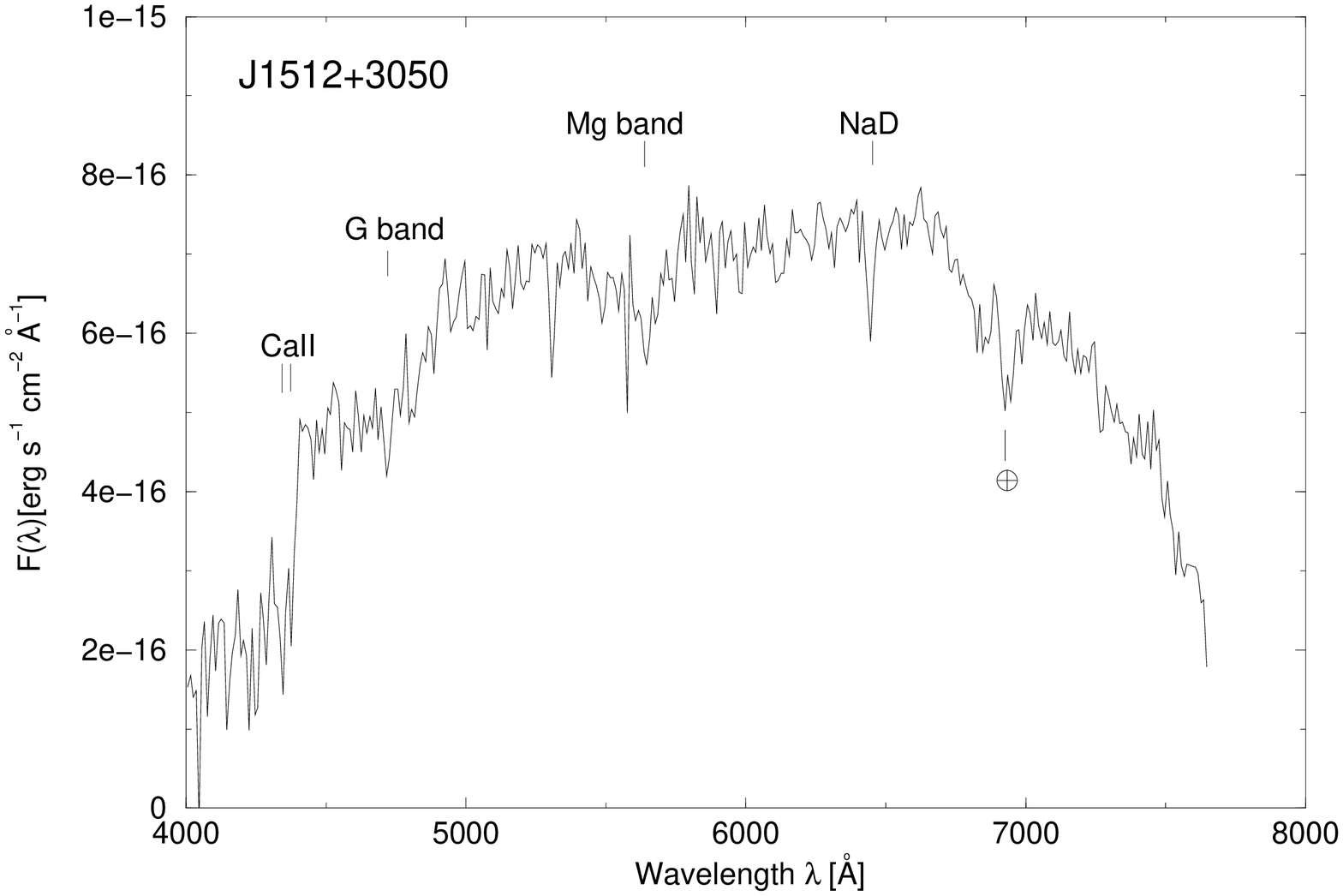}}
\caption{As in Fig.~4a but for the galaxy identified with the source J1512+3050}
\end{figure*}
 
\renewcommand{\thefigure}{4m}
\begin{figure*}
\resizebox{15cm}{!}{\includegraphics{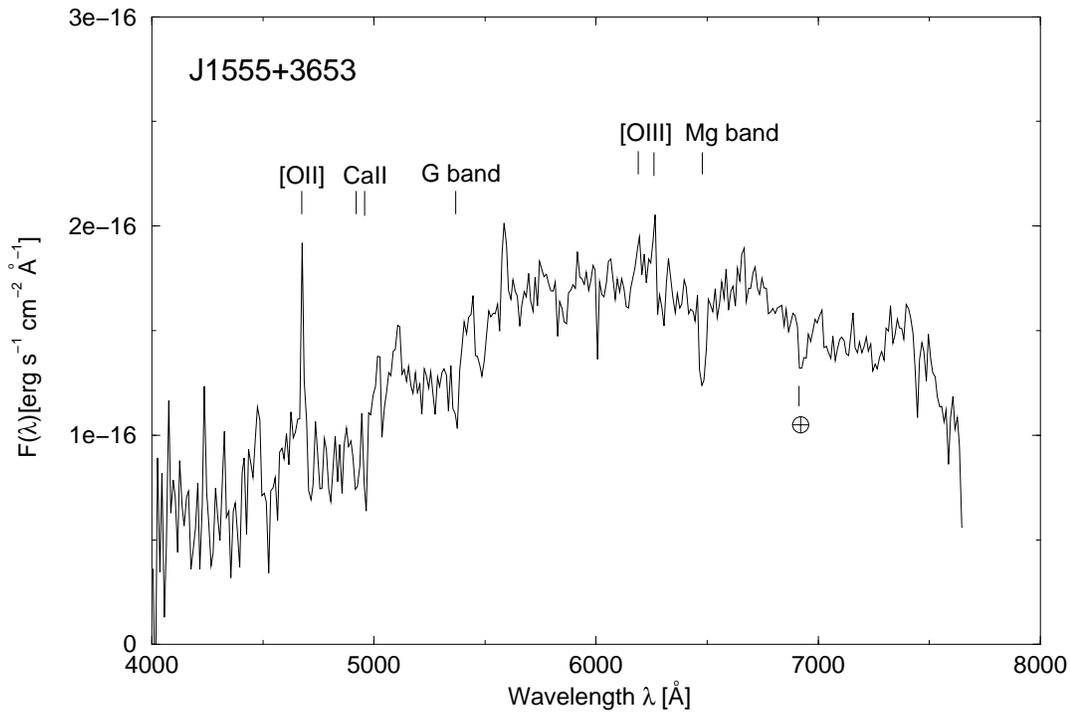}}
\caption{As in Fig.~4a but for the galaxy identified with the source J1555+3653}
\end{figure*}
 
\renewcommand{\thefigure}{4n}
\begin{figure*}
\resizebox{15cm}{!}{\includegraphics{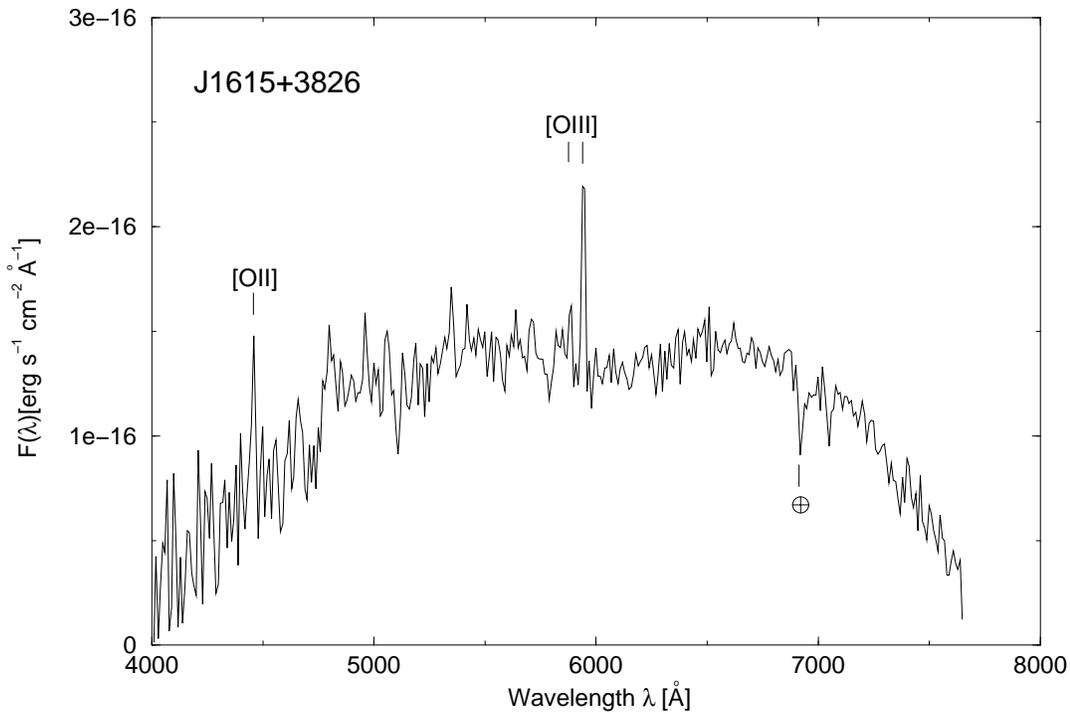}}
\caption{As in Fig.~4a but for the galaxy likely identified with the source
J1615+3826}
\end{figure*}
 
\renewcommand{\thefigure}{4o}
\begin{figure*}
\resizebox{15cm}{!}{\includegraphics{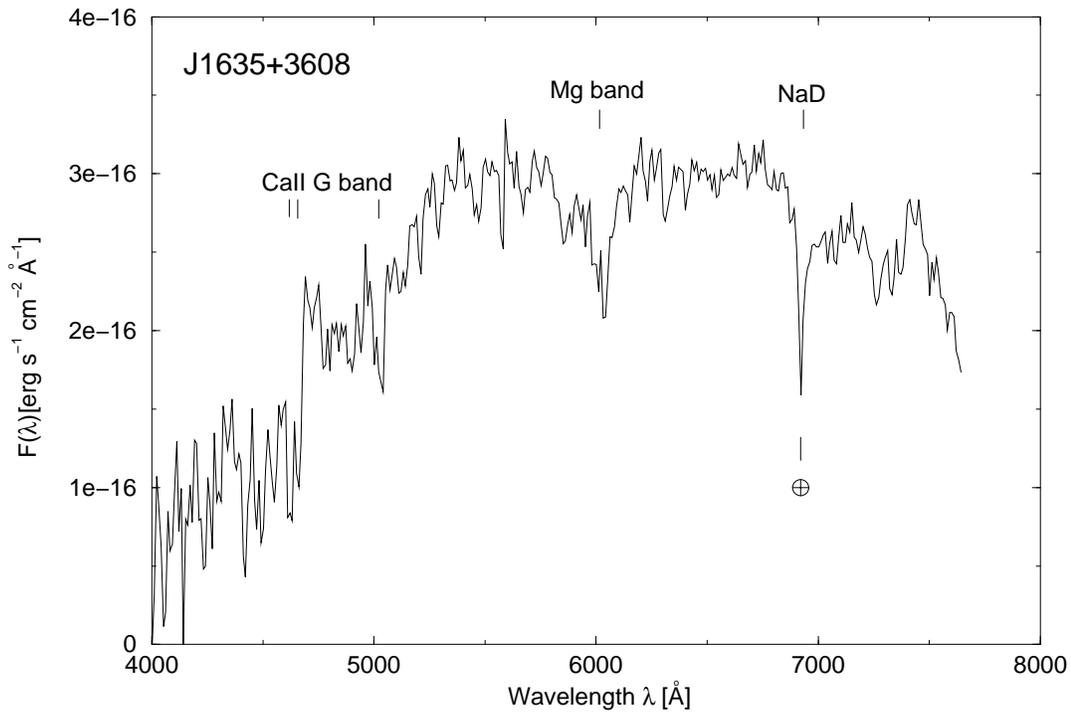}}
\caption{As in Fig.~4a but for the galaxy identified with the source J1635+3608}
\end{figure*}

\end{document}